\documentclass[3p,authoryear,preprint,review,times]{elsarticle}

\usepackage[T1]{fontenc}
\usepackage{amssymb}
\usepackage{amsmath}          
\usepackage{array}
\usepackage{adjustbox}
\usepackage{commath}
\usepackage{textcomp, gensymb}
\usepackage{caption}
\usepackage{multirow}
\usepackage{booktabs}
\usepackage{makecell}
\usepackage{subcaption}
\usepackage{bm}
\usepackage{natbib}
\usepackage{xurl} 
\usepackage[all]{nowidow}
\usepackage[hidelinks]{hyperref}

\usepackage[usenames, dvipsnames]{color}
\definecolor{myblue}{RGB}{0, 0, 205}
\definecolor{myred}{RGB}{205, 0, 0}
\definecolor{mypurple}{RGB}{155, 0, 155}
\definecolor{mygreen}{RGB}{0, 155, 0}
\definecolor{myturquoise}{RGB}{0, 155, 155}

\newcommand{\fig}[1]{Figure~\ref{#1}}
\newcommand{\tab}[1]{Table~\ref{#1}}
\newcommand{\eq}[1]{Equation~\ref{#1}}
\newcommand{\sect}[1]{Section~\ref{#1}}
\newcommand{\tmp}[1]{\textcolor{red}{[\bf #1]}}

\graphicspath{{figures/}}

\usepackage[switch]{lineno}

\journal{}

\begin{document}

\begin{frontmatter}


\title{Three-dimensional and clap-and-fling effects on a pair of flapping wings used for thrust generation}

\author[1]{Antoine Papillon\fnref{label4}}
\author[1]{Mathieu Olivier\corref{cor1}\fnref{label5}}
\fntext[label4]{antoine.papillon.1@ulaval.ca}
\fntext[label5]{mathieu.olivier@gmc.ulaval.ca}

\cortext[cor1]{Corresponding author}
\ead{mathieu.olivier@gmc.ulaval.ca}

\affiliation[1]{
organization={Department of Mechanical Engineering},
addressline={Universit\'e Laval},
city={Quebec City},
state={QC},
postcode={G1V~0A6},
country={Canada}
}

\begin{abstract}
This research investigates the three-dimensional effects of flow around a pair
of flapping wings undergoing clap-and-fling interactions through harmonic
in-plane motions combining pitching, heaving, and deviation. Both wings mirror
each other in their movements. Baseline parameters were selected based on
efficient configurations from a previous two-dimensional study. Simulations at a
Reynolds number of 800 were conducted on wings in a thrust configuration
utilizing the clap-and-fling mechanism. Rectangular flat-plate wings with
various aspect ratios were evaluated to assess the impact of three-dimensional
effects on performance. Results were compared with equivalent two-dimensional
computations.  It was found that the relative decrease in cycle-averaged
efficiency due to three-dimensional effects can be limited to approximately
5.59\% by using wings with an aspect ratio of 8, the largest considered in this
work. Moreover, wing proximity inherent to clap-and-fling kinematics helps
mitigate adverse three-dimensional effects.
\end{abstract}


\end{frontmatter}


\section{Introduction}
\label{Sec: intro}

Biomimetics is a field that has motivated numerous researchers to explore and
comprehend the physics governing the flight of various animal species, such as
birds and insects. The aim is to understand how these creatures generate their
lift or thrust during flight. One outcome of these investigations is the
development of systems such as Micro Air Vehicles (MAVs) that utilize flapping
wings. This approach contrasts with traditional intermittent flapping and
gliding observed with larger animals. Notably, a fascinating wing movement known
as clap-and-fling, initially discussed by \cite{weis-fogh_quick_1973}, has
naturally been adopted by certain birds and insects for flight. In this
mechanism, wings draw close to each other during flapping, fostering amplified
thrust or lift production due to their interaction. It has been shown by several
studies that the generated force benefits greatly from optimizing the wings'
kinematics as they get closer to each other during a cycle of oscillation
\citep{mao_flows_2003, zhu_effect_2018}.  Indeed, poor kinematics can impair the
flow due to the strong interaction between the wings, which may significantly
decrease aerodynamic performances. 

Similarly, earlier investigations into insect flight suggested that traditional
stationary aerodynamic principles fall short in predicting sufficient lift at
low Reynolds numbers necessary for hovering \citep{ansari_aerodynamic_2006,
dickinson_wing_1999, ellington_leading-edge_1996}. Instead, insects such as
fruit flies employ mechanisms such as the clap-and-fling mechanism in such
scenarios, creating circulation around each wing through mutual interaction. The
circulation of each wing serves as the starting vortex of the other one, which
minimizes the time required to reach a steady state.  Consequently, lift
increases during each oscillation cycle \citep{lighthill_weis-fogh_1973}.
This explains why several species adopted the clap-and-fling mechanism at lower
Reynolds numbers, even though some species don't use it
\citep{dickinson_wing_1999}.

Another study, conducted via numerical simulations, compared the outcomes of a
clap-fling-sweep motion for hovering by analyzing both two-dimensional (2D) and
three-dimensional (3D) Computational Fluid Dynamics (CFD) simulations.
\cite{kolomenskiy_vorticity_2010} first obtained 2D results for one and two
flapping wings using a Fourier pseudo-spectral method with volume penalization.
One of the main takeaways from this study is that leading-edge vortices increase
the circulation and lift of the wings when they remain close to the latter. For
a single flapping wing, this is counteracted by an upward flow past the trailing
edge. However, in the case of two flapping wings, the interaction between the
two bodies creates a pressure jump across the hinge, suppressing the detrimental
upward flow. The subsequent study, which included 3D computations
\citep{kolomenskiy_two-_2011}, was based on these conclusions. In the
clap-fling-sweep motion, the wings initially clap as their upper surfaces touch,
then fling open like a book, creating circulation corresponding to the starting
vortex of each wing. Lastly, the wings sweep apart symmetrically, generating
lift by carrying bound vortices. The study indicated that, while 2D simulations
sufficiently approximated lift during the fling phase, 3D effects become
significant during the sweep motion, delaying stall and enhancing lift
production \citep{kolomenskiy_two-_2011}.

Previously, it was established that most insects and small birds favor the
clap-and-fling mechanism at lower Reynolds numbers. However, investigations at
higher Reynolds numbers were conducted to assess force generation using this
motion. \cite{phan_clap-and-fling_2016} conducted 3D numerical computations and
experimental measurements on a hovering two-winged flapping-wing micro air
vehicle at a Reynolds number of 15,000. The results revealed a 6\% disparity
between the measured forces of the prototype and the CFD model, validating the
numerical outcomes. The clap-and-fling mechanism increased the average vertical
force by 11.5\%, and the horizontal drag force by 18.4\%, though overall
efficiency declined at this high Reynolds number. Furthermore, the study
demonstrated that the mechanism's impact becomes negligible when the minimum
distance between the two wings exceeds 1.2 times the chord length, emphasizing
the importance of wing proximity.

Another 3D study by \cite{cheng_aerodynamic_2017} analyzed two variations of the
clap-and-fling mechanism at a Reynolds number of 20, relevant to small flying
insects. The full configuration involved the wings clapping close together along
their span, producing two force peaks 8–9 times larger than for a single-wing
case, occurring at the end of the clap and the beginning of the fling phase.
However, these peaks primarily generated drag, with minimal lift enhancement,
reducing efficiency. In the partial configuration, only the wing tips clapped
closely, with separation increasing toward the wing roots. This reduced wing
interaction prevented significant drag forces while increasing mean lift by 12\%
without notable efficiency deterioration, making it more suitable for small
insects.

Flexible wings have also been investigated in several 3D numerical studies.
\cite{noyon_effect_2014} used an immersed boundary method to study flapping
wings with active chord-wise deformation, demonstrating that increasing the
leading-edge angle of attack accelerates vortex formation, boosting lift.
\cite{tay_numerical_2018} simulated the clap-and-fling motion with wing
interactions, achieving reasonable agreement between CFD and particle image
velocimetry experiments. For higher Reynolds numbers (8,875), drag forces were
about 2.5 times greater than for single-wing cases, consistent with the findings
by \cite{miller_computational_2005}, which showed that clap-and-fling motions at
very low Reynolds numbers (8–128) produce drag forces up to 10 times higher than
those without wing-wing interactions. Flexibility, however, reduced drag and
improved efficiency.  Similarly, \cite{jadhav_effect_2019} analyzed passive wing
flexibility on a micro air vehicle. They showed that minimizing wing separation
during clapping enhances leading-edge vortices and total lift.  However, passive
flexibility had a negligible effect on lift compared to rigid wings, suggesting
that careful adjustments to wing distance are essential for optimizing hovering
configurations.

Most previous studies have focused on delving into the aerodynamic performance
of flapping wings in a hovering flight setup, probing the effects of various
parameters. Recent work by the authors \citep{Papillon_clap_fling_2024} expanded
upon these insights using 2D numerical simulations carried out at a Reynolds
number of 800, particularly investigating the utilization of flapping wings for
thrust generation through the clap-and-fling and deviation mechanisms. A
propulsive efficiency of 0.560 was obtained, an increase of 17.14\% with respect
to an equivalent single-foil configuration.  This improvement involved
fine-tuning the phase and amplitude of the pitching and deviation motions in
configurations involving near contact of the two foils.  For a fixed pivot point
positioned at the foils' leading edge, the study revealed that an optimal phase
shift of $105\degree$ between the heaving and pitching motions could notably
enhance the efficiency and thrust coefficient. This finding corroborates the
earlier conclusions by \cite{wang_unsteady_2004}, emphasizing the sensitivity of
force coefficients to the phase relationship between heaving and pitching
motions.  This optimized phase shift results in the minimum spacing occurring at
the foils' trailing edges during the motion, which leads to significant
phenomena throughout the motion cycle. During the clap phase, a high-pressure
zone contributing positively to thrust appeared to the right of the foils as
they approached each other at their leading edges with a spacing slightly larger
than the subsequent minimum spacing occurring at their trailing edges. As the
foils' inclination changes, fluid between them is swiftly ejected downstream
when they reach this minimum spacing at the trailing edges.  Subsequently, a
low-pressure zone emerges in front of the wings during the ensuing fling phase
due to the generation of robust leading-edge vortices.  The study also showed
that the foils must reach near contact during flapping to leverage the
clap-and-fling phenomenon. 

Moreover, the deviation motion, which is essentially a harmonic translation of
the wings parallel to the flow, is based on real fruit flies wing kinematics
which employ a slightly more complex motion to benefit from the clap-and-fling
phenomenon \citep{fry_aerodynamics_2003}. Results showed that by adjusting the
preponderance of deviation through its amplitude at a moderate value and by
adjusting the phase shift between the deviation and the heaving motion, the
unsteady effects of the clap-and-fling mechanism can be enhanced by modulating
the effective velocity and angle of attack of the foils at key instants in the
cycle. Additionally, it was shown that at higher positive phase shift values,
increasing the deviation amplitude has the general trend of leveling the
aerodynamic forces of the system, which was consistent with the findings of
\cite{bos_influence_2008}.

This paper aims to expand on these foundations by investigating the
three-dimensional effects on a pair of flapping wings subjected to the
clap-and-fling mechanism through numerical simulations. Our research focuses on
the aerodynamic performance of flapping wings used for propulsion, which
contrasts with the majority of anterior studies which considered hovering flight
configurations. The pair of flapping wings undergo a combined heaving, pitching,
and deviation motion based on the best configuration obtained in the previous
work presented by \cite{Papillon_clap_fling_2024}. Such an optimal
configuration, with near-contact distances between the foils, has been shown to
significantly increase thrust and efficiency. Hence, this paper aims to build
upon the previous results and extend their scope to finite-span 3D wings to
analyze how this more realistic geometry affects the surrounding flow and thus,
the performances of the propulsive system. To properly handle the motion of the
wings, the overset-mesh technique is utilized between each mesh region which
allows proper interpolation. Then, simulations with wings of various aspect
ratios (span to chord length) are executed to assess three-dimensional effects
on the performance metrics. Additionally, an adaptive mesh refinement technique
based on vorticity is used in the numerical simulations to reduce computational
costs, while maintaining proper accuracy. The paper's structure is as follows:
\sect{Sec: methodology_2} outlines the numerical methodology and performance
metrics used to characterize the propulsive system. \sect{Sec: verification_2}
verifies the numerical simulations, while \sect{Sec: results_2} presents the
numerical study outcomes about the three-dimensional effects on the aerodynamic
performances.

\section{Methodology}
\label{Sec: methodology_2}

\subsection{Problem definition}
\label{Sec: dynamics_2}

\begin{figure}[!ht]
\centering
\includegraphics[width=0.8\columnwidth]{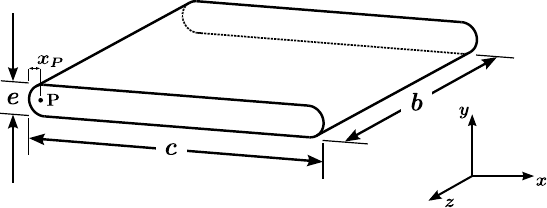}
\caption{Geometry of a thin flat-plate wing with rounded edges.} 
\label{Fig_geometry_plate_2}
\end{figure}

\begin{figure}[!ht]
\centering
\includegraphics[width=0.65\columnwidth]{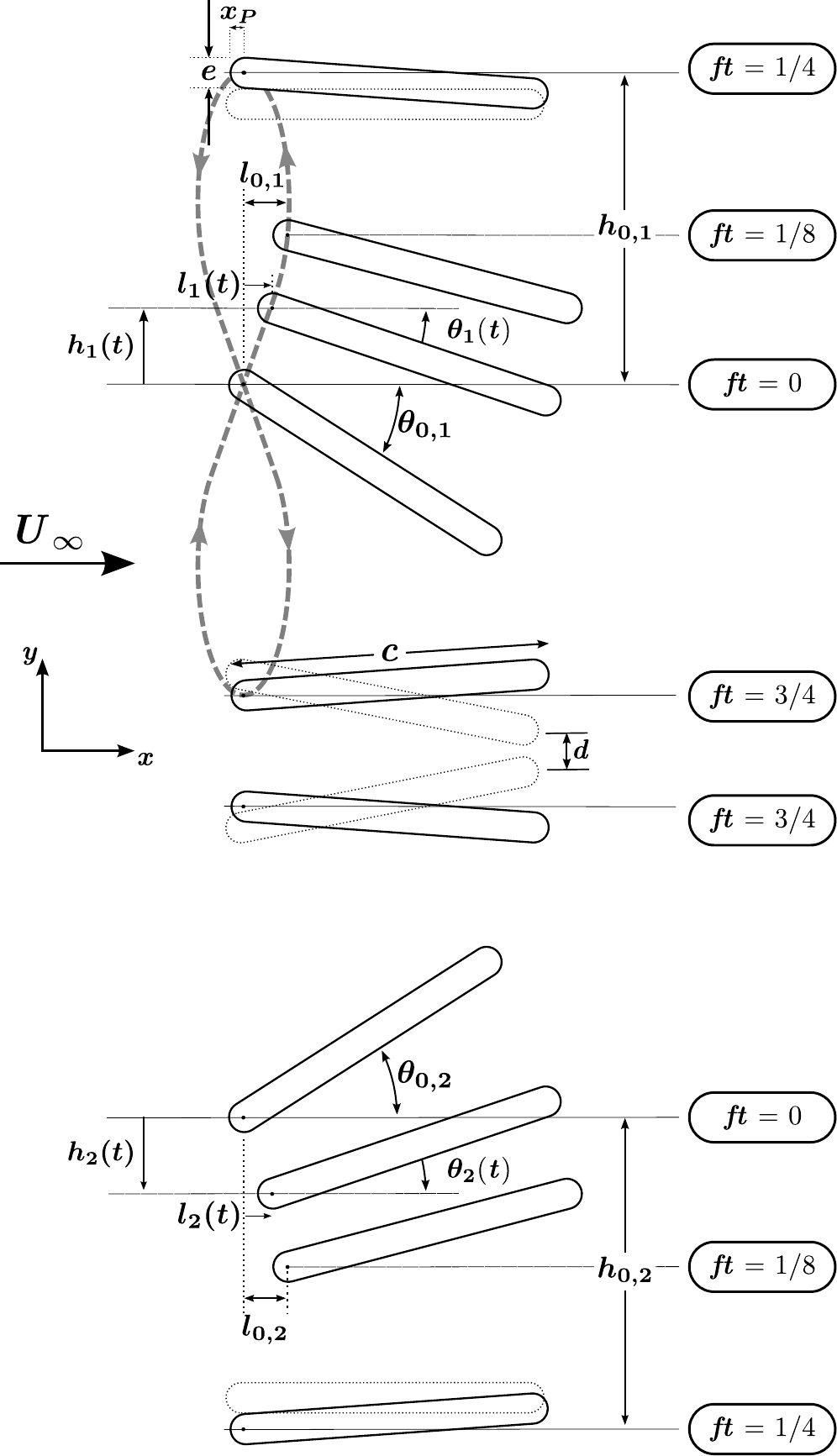}
\caption{In-plane motion of the two flapping wings.}
\label{Motion_two_plates_2}
\end{figure}

Taking inspiration from natural flapping-wing flyers at low Reynolds numbers, we
used a pair of wings shaped as thin flat plates as illustrated in
\fig{Fig_geometry_plate_2} for the numerical simulations. The parameters
describing wings are the chord length $c$, the thickness $e$, and the span $b$,
whereas $x_P$ is the pivot point distance from the leading edge.
The motion of each flapping wing comprises three components: vertical
translation (heaving motion), rotation (pitching motion), and horizontal
translation (deviation motion). Such a motion is equivalent to the simplified
one used in two-dimensional cases \citep{Papillon_clap_fling_2024}, which allows
a direct analysis of the three-dimensional effects.  Moreover, while this
in-plane motion is not the same as that of flying species, it remains
interesting from a mechanical design perspective.  For this study, each motion
component is thus respectively described by:  
\begin{equation}
h = h_0 \sin\left(2 \pi ft \right),
\label{eq_harmonic_1_2}
\end{equation}
\begin{equation}
\theta = \theta_0 \sin\left(2 \pi ft - \phi \right),
\label{eq_harmonic_2_2}
\end{equation}
\begin{equation}
l = l_0 \sin\left(4 \pi ft - \psi \right),
\label{eq_harmonic_3_2}
\end{equation}
where the amplitudes of heaving, pitching, and deviation motions are
respectively $h_0$, $\theta_0$, and $l_0$, the motion frequency is represented
by $f$, and $\phi$ and $\psi$ are phase parameters. Also, the amplitudes $h_0$
and $\theta_0$ of the two wings have opposite signs in the coordinate system
illustrated in \fig{Fig_geometry_plate_2}. The in-plane motion of the two
flapping wings is illustrated in \fig{Motion_two_plates_2}. In this figure, the
distance $d$ represents the minimum spacing reached between the wings throughout
the motion cycle, and the flow velocity, denoted as $U_{\infty}$, corresponds to
the baseline horizontal velocity of the system without considering the effects
of the deviation motion.

\subsection{Parametric Space}
\label{Sec: parametric_s_2}

The main reference scales of the problem are: the length scale $c$, velocity
scale $U_{\infty}$, time scale $1/f$, and pressure scale $\rho U_\infty^2$.
These scales result in the following dimensionless parameters:
\begin{equation}
\begin{split}
& Re=\frac{U_\infty \, c}{\nu}, \quad
f^*=\frac{f \, c}{U_\infty}, \quad
h^*=\frac{h_0}{c}, \quad
l^*=\frac{l_0}{c}, \quad \\
& e^*=\frac{e}{c}, \quad
x_P^*=\frac{x_P}{c}, \quad 
d^*=\frac{d}{c}, \quad
b^*=\frac{b}{c}, \quad
\theta_0, \quad
\phi, \quad
\psi.
\label{eq_parametric_space_2}
\end{split}
\end{equation}
These parameters are respectively the Reynolds number, the reduced frequency,
the normalized heaving amplitude, the normalized deviation amplitude, the
relative thickness of the airfoil, the normalized pivot point position, the
normalized minimum spacing between the wings, the normalized span, the pitching
amplitude, the phase shift between heaving and pitching, and the phase shift
between heaving and deviation. Since the focus is on varying the normalized span
to characterize the impact of three-dimensional effects, the following
parameters are used throughout the study:
\begin{equation}
\begin{split}
& Re=800, \quad h^*= 1, \quad l^*= 0.035,
\quad e^*=0.01, \quad x_P^*=0.005, \\
& \theta_0=30\degree, \quad f^*=0.15, \quad d^*=0.005,
\quad \phi=105 \degree, \quad \psi=-125 \degree.
\label{eq_parametric_values_2}
\end{split}
\end{equation}

Most of these parameters are the same as in our previous work
\citep{Papillon_clap_fling_2024} and are representative of typical flying
species such as insects and small birds \citep{azuma_biokinetics_2006,
tobalske_three-dimensional_2007, weis-fogh_quick_1973}. The amplitude $h_0$ was
selected by considering the projection of the arc length of a 3D motion inspired
by insects at a fraction of the wing span onto a 2D plane
\citep{miller_computational_2005, wang_unsteady_2004}. The pivot point position
draws inspiration from birds' wings driven by forelimbs. In the context of
in-plane motion, it corresponds to the leading edge of the wings. The parameters
$f^*$ and $\theta_0$ are those achieving optimal efficiency for a single wing at
a phase parameter of $\phi = 90\degree$ \citep{olivier_parametric_2016}.
However, our recent work \citep{Papillon_clap_fling_2024} established that in
the case of two flapping wings undergoing the clap-and-fling phenomenon, the
most efficient configuration occurs at $\phi = 105\degree$ along with a
deviation characterized by $l^*=0.035$ and a phase shift $\psi=-125 \degree$.
The normalized distance $d^*$ of 0.005 is used since it increases both thrust
and propulsive efficiency. While our previous results suggest that a slight
performance improvement might still be obtained with smaller values of $d^*$,
doing so would put additional resolution constraints on the numerical method.
Lastly, since the scope of this paper is to tackle 3D effects, the main
parameter of the study is the normalized span $b/c$.

\subsection{Performance metrics}
\label{Sec: perf_metrics_2}

The flapping-wing configuration depicted in \fig{Motion_two_plates_2} undergoes
a horizontal deviation motion at a velocity $\dot{l}$. Considering the
horizontal flow velocity $U_{\infty}$, the effective horizontal velocity is thus
$U_{\infty}-\dot{l}$. When the wings translate leftward, their effective
velocity exceeds the relative velocity $U_{\infty}$, while the opposite holds
for rightward translation. Given the suggested orientation for the $x$ and $y$
axes, the net thrust force, accounting for the force necessary for deviation
motion, is denoted as $-F_{x}$. Meanwhile, $F_{y}$ and $M_{z}$ contribute to
sustaining the flapping motion of each wing. The instantaneous thrust
coefficient for wing $i$ is defined as follows:
\begin{equation}
    C_{T,i} = - \frac{F_{x,i}}{\frac{1}{2} \, \rho_f \, U_\infty^2 \, c \, b}.
\label{Eq_Ct}
\end{equation}
Similarly, the power coefficient required to sustain the motion of wing $i$ is
defined as:
\begin{equation}
    C_{P,i} = -\frac{F_{y,i}\, \dot{h_i} \ + \ M_{z,i} \, \dot{\theta_i}
    \ + \ F_{x,i}\, \dot{l_i}}{\frac{1}{2} \, \rho_f \, U_\infty^3 \, c \, b}.
\label{Eq_Cp}
\end{equation}
These two equations assume the two wings are part of a system that moves at
constant velocity $U_{\infty}$. Therefore, the power required to activate the
deviation motion $F_{x,i}\, \dot{l_i}$ is added to the total power consumption.
This power coefficient formulation is thus equivalent to that of
\cite{Stroke_deviation_2013}. To accommodate the fluctuating behavior of these
force and power coefficients, time-averaging denoted by the operator
$\overline{()}$ is applied once the flow attains a periodic state, usually
observed after 3-5 cycles. This averaging process spans the last cycle of
motion. The overall average coefficients, encompassing contributions from both
wings, are computed as follows:
\begin{equation}
    \overline{C_T} =  \frac{\overline{C_{T,1}} + \overline{C_{T,2}}}{2},
\label{Eq_Ct_glob_2}
\end{equation}
\begin{equation}
    \overline{C_P} =  \frac{\overline{C_{P,1}} + \overline{C_{P,2}}}{2}.
\label{Eq_Cp_glob_2}
\end{equation}

Another important metric is the thrust efficiency of the flapping wings,
indicating the relationship between the thrust power $\left(F_x \,
U_{\infty}\right)$ and the power necessary for executing heaving, pitching, and
deviation motions. This efficiency is defined for a single wing as:
\begin{equation}
    \eta_i = \frac{\overline{F_{x,i} \, U_{\infty}}}
    {\overline{F_{y,i}\, \dot{h_i} \ + \ M_{P,i} \, \dot{\theta_i} \ + \ F_{x,i}\, \dot{l_i}}}
    = \frac{\overline{C_T}}{\overline{C_P}}.
\label{Eq_eta}
\end{equation}
The total average efficiency, taking into account both wings, is then expressed
as:
\begin{equation}
    \eta =  \frac{\eta_1 + \eta_2}{2}.
\label{Eq_eta_glob}
\end{equation}


\subsection{Computational domain and numerical details}
\label{Sec: domain}

\begin{figure}[!ht]
\centering
\includegraphics[width=0.8\textwidth]{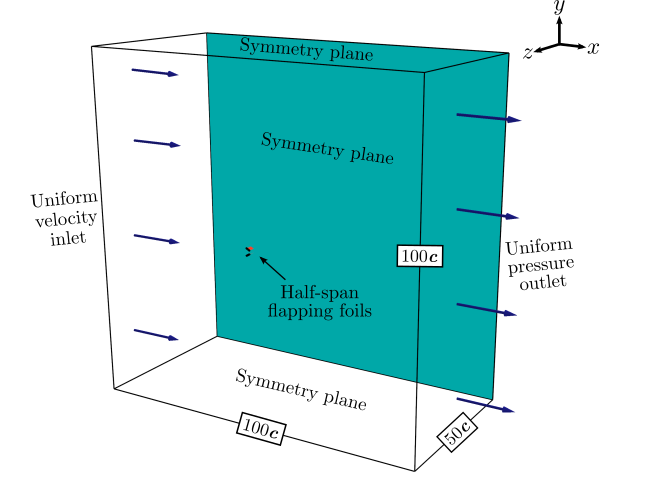}
\caption{Computational domain and boundary conditions used for the simulations 
of two flapping wings.}
\label{domain_2}
\end{figure}

\fig{domain_2} depicts the computational domain utilized in this study—a $100c
\times 100c \times 50c$ box. The two wings are initially positioned at a
distance neighboring the chord length $c$ from the horizontal axis at the domain
center and ensure the minimum spacing $d^*$ according to \eq{eq_harmonic_1_2}.
The wings are also placed at an initial horizontal position and pitching angle
derived from \eq{eq_harmonic_2_2} and \eq{eq_harmonic_3_2} relative to the
center of the domain. Boundary conditions are as follows: symmetry conditions
are imposed on the top, bottom, and back domain sides, a uniform velocity inlet
condition is imposed on the left side, and a pressure outlet boundary condition
with uniform static pressure is set on the right side. The 3D simulations take
advantage of the symmetry by modeling only half the domain through the use of
the front symmetry plane, hence the length of the computational box of $50c$ in
the spanwise direction. The incompressible Navier-Stokes equations govern the
flow field within the domain. The solution is computed with STAR-CCM+ which
employs a second-order accurate finite-volume method supporting unstructured
polyhedral overset meshes. Temporal advancement relies on a second-order
backward difference scheme. 

\begin{figure*}[!ht]
\centering
\begin{subfigure}[t]{.49\textwidth}
\centering
\includegraphics[width=\linewidth]{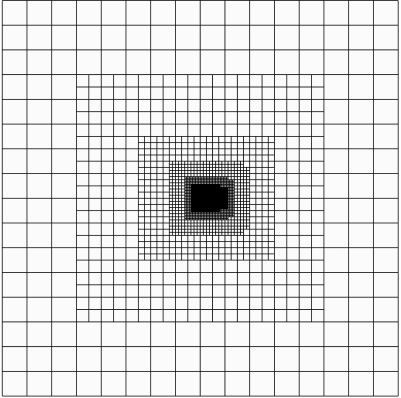}
\caption{Background mesh used in the simulations.}
\label{mesh_ambiant}
\end{subfigure}
\hfill
\begin{subfigure}[t]{.49\textwidth}
\centering
\includegraphics[width=\linewidth]{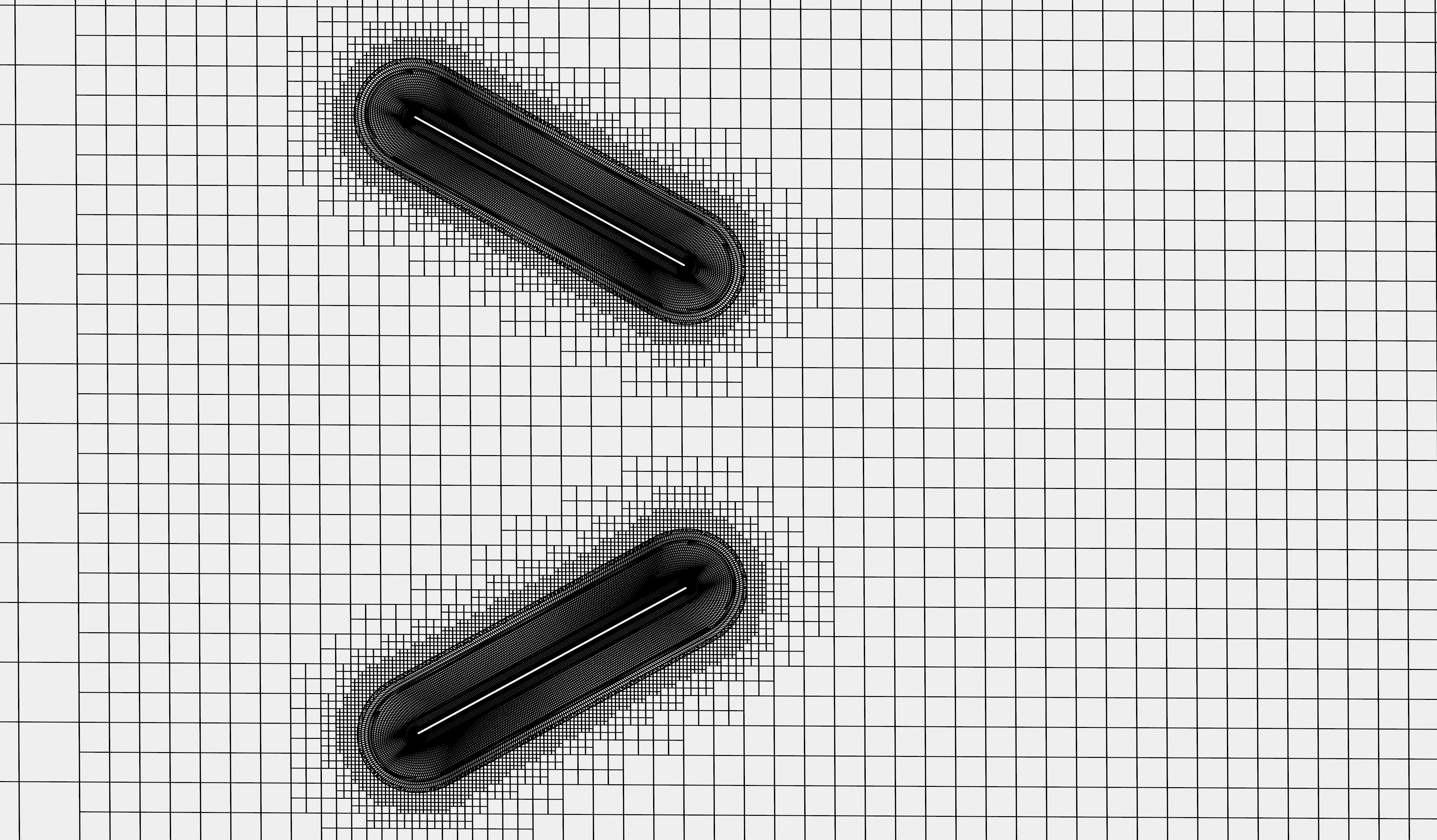}
\caption{Adaptive mesh refinement region around the two wings.}
\label{mesh_AMR_zone}
\end{subfigure}

\begin{subfigure}[t]{.85\textwidth}
\centering
\includegraphics[width=\linewidth]{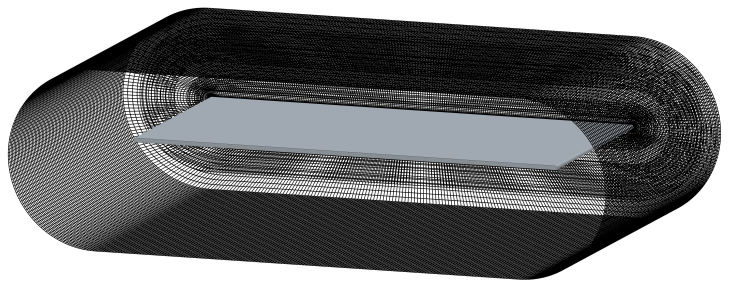}
\caption{Overset mesh around a single wing.}
\label{mesh_plate}
\end{subfigure}
\caption{Mesh details.}
\label{mesh}
\end{figure*}

The mesh employed for the presented simulations (\fig{mesh}) comprises a
background grid constructed with the trimmed-cell mesh generator. An increased
resolution is used in the central region encompassing the two wings
(\fig{mesh_ambiant}). The background mesh is initially coarse, and it uses an
adaptive mesh refinement technique to match the resolution of the overset meshes
around the wings and to refine accordingly in the near wake region
(\fig{mesh_AMR_zone}). Indeed, each wing possesses an individual mesh set up as
an overset region to perform the clap-and-fling motion.  These overset meshes
were constructed with the ICEM-CFD software, which provides better control over
mesh element distributions. These structured meshes also allow the definition of
hierarchical regions to control which cells the interpolation algorithm uses
when the two overset-mesh regions overlap during the clap-and-fling phase.


\section{Numerical verifications}
\label{Sec: verification_2}

\subsection{Iterative convergence}
\label{Sec: convergence}

The flow solver utilizes the SIMPLEC iterative algorithm to manage the coupling
between pressure and velocity. Achieving accurate results requires meeting
specific stopping criteria at each time step to prevent significant iterative
errors.  Convergence of the iterations is assessed through an asymptotic
verification of six key physical quantities: the $x$-force, $y$-force,
$z$-force, and three aerodynamic moments ($M_{x,i}, M_{y,i}, M_{z,i}$) computed
for each wing. With simulations involving two wings, twelve stopping criteria
are thus applied. These criteria set an asymptotic limit of $10^{-4}$,
normalized using the averaged solution obtained from five samples.  Because
these criteria directly tackle the forces and moments used in calculating the
performance metrics outlined in \sect{Sec: perf_metrics_2}, the effect of the
iterative error on the performances is directly controlled. Moreover, it was
noted that the residuals of all equations consistently decreased at each time
step in all simulations. 

\subsection{Adaptive mesh refinement setup}
\label{Sec: AMR}

The numerical simulations performed in this study are three-dimensional, which
entails high computational time. Thus, an adaptive mesh refinement technique is
employed to save computational time. As mentioned, a box inside the background
grid with coarser cells is set up around the two overset regions to allow
adaptive mesh refinement for proper interpolation according to specific
parameters. Indeed, the overset mesh refinement is configured with a maximum of
three refinement levels in a way where the finer cells are of similar size to
the cells on the external boundaries of the overset regions of the wings. This
can be observed in \fig{mesh_AMR_zone} which shows a two-dimensional mesh
section. The overset mesh refinement is also performed in the spanwise direction
with the refinement box exceeding $0.5c$ the foils' tips. 

Additionally, the adaptive mesh refinement is utilized for wake refinement
purposes. The solver needs a criterion to either coarsen or refine a mesh
element in the relevant region. In this problem, the wings perform the
clap-and-fling motion and periodically eject vortices in the wake as they flap.
Consequently, mesh adaptation is relevant for tracking the vortices as they
travel downstream. The normalized vorticity-based criterion used to determine
whether the mesh should be refined or coarsened is given by $\abs{\nabla
\abs{\vec{\omega}} \cdot c \, \Delta x \, / \, U_\infty}$, where $\vec{\omega}$
is the vorticity vector. This criterion uses the gradient of the vorticity
magnitude to measure changes in vorticity.  Consequently, mesh refinement and
coarsening are based on a metric estimating the normalized vorticity difference
between neighboring cells. The magnitude of this product is then performed and
utilized as the final criterion in the simulations. Moreover, a threshold has to
be fixed to determine when to refine, coarsen, or keep a mesh element. The range
used in this study, which yielded an adequate mesh refinement in the wake, is
$[0.15, 0.75]$. When the value of the criterion exceeds the specified maximum,
the cell is refined.  Similarly, the mesh is coarsened if the value goes below
the minimum and the element size is kept when it is between the two specified
values. Moreover, the adaptive mesh refinement is also activated near the
overset regions of the wings to bridge the cell size gap between the background
grid and the latter.  Three maximum levels of refinement are globally specified
to ensure a proper transition and precise computations. The mesh refinement is
done by splitting the cells along the three directions. This is performed every
10 time steps to control the interpolation error while updating frequently
enough to capture the near wake and the ejected vortices.  

\begin{figure}[!ht]
\centering
\includegraphics[width=\columnwidth]{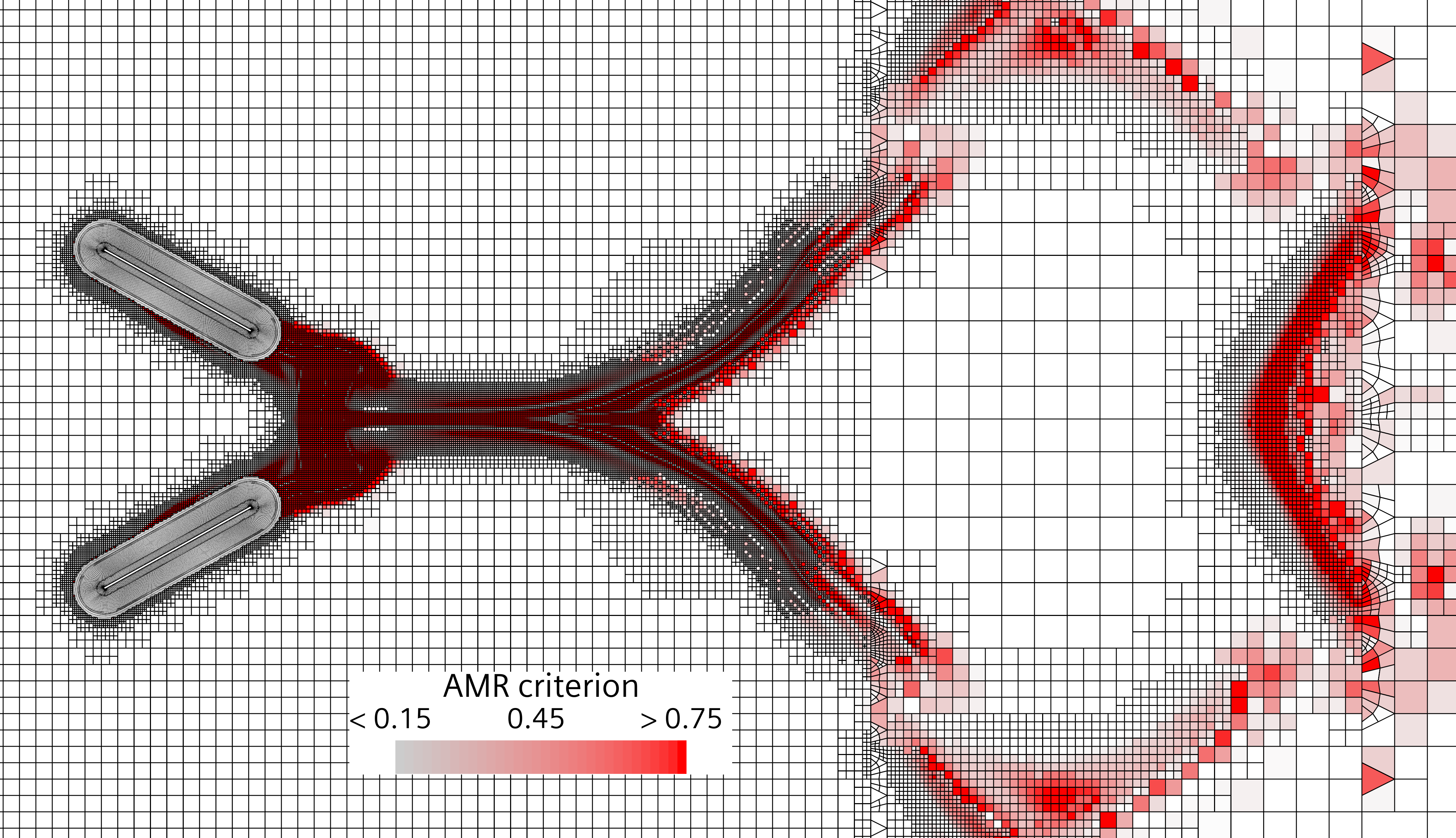}
\caption{Adaptive mesh refinement in the wake for the case with a normalized
span of $b^*=1$ (at the instant $ft=0$).}
\label{fig_AMR_vorticity}
\end{figure}

A section of the mesh showing the refinement performed in the wake can be
observed in \fig{fig_AMR_vorticity}.
It depicts the wake refinement performed for the case where the wings have a
normalized span of $b^*=1$. This illustration occurs at the beginning of a
motion cycle after several oscillation periods, hence the previously shed
vortices are visible in the wake through the mesh refinement.  The scalar field
of the expression for the refinement criterion is also superimposed to the mesh
to observe where the changes in vorticity are the highest. As seen, those
changes are the greatest near the overset regions of the wings and then separate
into two thin bands downstream. The cells are thus refined across multiple
levels. Outside of those regions, the changes in vorticity are lower than the
specified minimum and the mesh stays in its default coarse state. Thus, while
the refinement metric is arbitrary, the resulting refined mesh is consistent
with the physics of the problem.

\subsection{Mesh and time step convergence}
\label{Sec: independence}
\begin{figure}[htb]
\centering
\includegraphics[width=\linewidth]{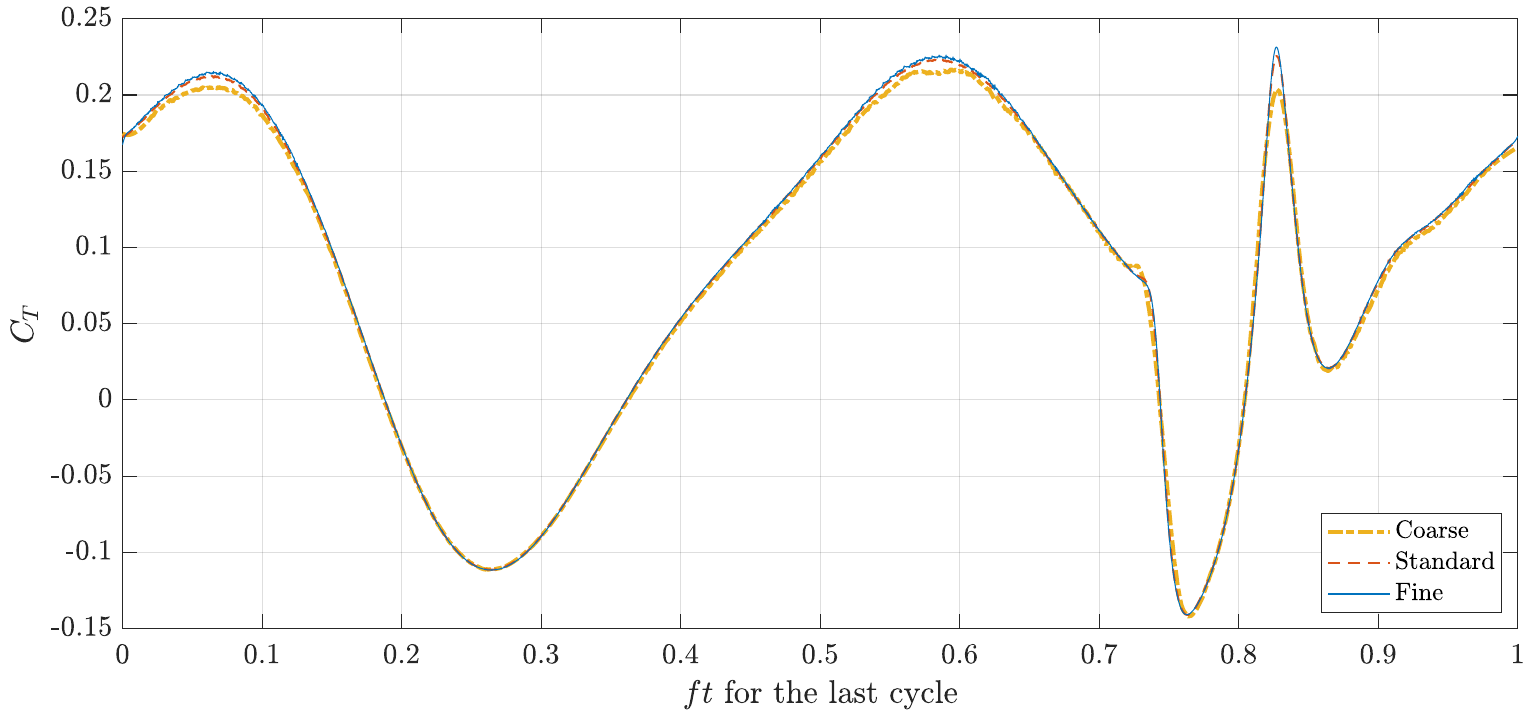}
\caption{Convergence for the thrust coefficient for the last cycle of motion.}
\label{thrust_ev_glob}
\end{figure}

Three distinct meshes were used to verify and evaluate the discretization error:
a baseline mesh, a coarser version, and a finer version.
\tab{Tbl_verification} provides the cell counts for each mesh. To maintain
consistency, three distinct time-step values were employed, ensuring a constant
Courant number. The baseline time step was $f \Delta t = 1/2000$.
The adaptive mesh refinement algorithm was used for the three meshes considered and was configured to perform up to three refinement levels. \fig{thrust_ev_glob} shows the
instantaneous thrust coefficient during the final motion cycle for the
three meshes and their respective time steps. The motion parameters in these simulations
correspond to the values outlined in \sect{Sec: parametric_s_2}. Hence, the
wings' trailing edges approach each other closely at the end of the clap
phase. For the verification process, wings with a normalized span of $b^*=1$
were used which helped reduce the computational costs required to compute the
simulation using the finer mesh and time step.
While some differences remain visible around the peaks and troughs of thrust,
particularly for the curve associated with the coarser mesh, similarities in the
curves in \fig{thrust_ev_glob} suggest strong numerical convergence. The
observations from \tab{Tbl_verification} support these observations by assessing
the performance metrics introduced in \sect{Sec: perf_metrics_2}. The results
indicate that concurrent refinement in spatial and temporal domains has minimal
impact on the reported physical parameters in the table, thereby confirming the
appropriateness of the chosen baseline mesh and time step. The cell count
reported in \tab{Tbl_verification} is obtained at the beginning of a cycle
($ft=0$) after several periods of motion. Indeed, due to the motion of the wings
and the adaptive mesh refinement technique related to the ejection of vortices
in the wake in previous cycles, the number of cells in the background region and
thus, the total cell count, inherently changes at each time step.
\begin{table}[htb]
\caption{Convergence study summary: discretization details and cycle-averaged thrust and power coefficients.}
\begin{center}
\label{Tbl_verification}
\begin{tabular}{lccccc}
\toprule

Description &
Total cell count$^1$ &
\makecell{Overset zone \\ cell count} &
$\Delta t \cdot f$  &
$\overline{C_T}$ & $\overline{C_P}$ \\

\midrule
Coarse Mesh   & 1~089~484 & 332~150 & 1/1000 & 0.0826 & 0.381 \\
Baseline Mesh & 7~208~537 & 2~548~750 & 1/2000 & 0.0848 & 0.384 \\
Fine Mesh     & 40~564~128 & 15~762~296 & 1/4000 & 0.0855 & 0.385 \\
\bottomrule
\multicolumn{6}{l}{\footnotesize $^1$ The total cell count values were obtained at $ft = 0$ after several periods of motion.} \\
\end{tabular}
\end{center}
\end{table}

As in our previous study \citep{Papillon_clap_fling_2024}, we evaluated the
accuracy of the simulations when the leading and trailing edges of the wings
draw very close, i.e., when the two overset-mesh regions overlap. In
\fig{thrust_ev_glob}, a thrust peak occurs around $ft=0.825$, coinciding with
substantial overlap between the overset meshes of each wing. This prompts
concerns regarding the cell count in the gap between the wings, as the so-called
hole-cutting procedure used by the algorithm to select the mesh cells used for
the overset mesh interpolation affects mesh resolution and interpolation quality
in that region. To address this concern, \tab{Tbl_verification_loc} presents
local averages for power and thrust coefficients during the final cycle of each
simulation within the normalized period $ft \in [0.80,\ 0.85]$. This range
corresponds to the thrust peak where the trailing edges reach their minimum
spacing $d^*$, followed by the subsequent fling of the wings. The differences in
thrust and power coefficients, although slightly higher than those for
complete-cycle averages, remain acceptably low, indicating an adequate level of
convergence even when the wings are in proximity. These verifications confirm
that the baseline mesh and time step are suitable for the parametric study. In
cases with larger normalized spans $b^*$, the cell count naturally increases
since the resolution is preserved along the span.
\begin{table}[htb]
\caption{Verification of the discretization in the time interval  $ft\in [0.725,\ 0.850]$.}
\begin{center}
\label{Tbl_verification_loc}
\begin{tabular}{lccccc}
\toprule
Description &
Total cell count &
\makecell{Overset zone \\ cell count} &
$\Delta t \cdot f$  &
$\overline{C_T}\,^1$ & $\overline{C_P}\,^1$ \\

\midrule
Coarse Mesh   & 1~089~484 & 332~150 & 1/1000 & 0.114 & 0.301 \\
Baseline Mesh & 7~208~537 & 2~548~750 & 1/2000 & 0.115 & 0.301 \\
Fine Mesh     & 40~564~128 & 15~762~296 & 1/4000 & 0.115 & 0.300 \\
\bottomrule
\multicolumn{6}{l}{\footnotesize $^1$ $\overline{C_T}$ and $\overline{C_P}$ values were averaged on the normalized time interval $[0.725,\ 0.850]$.} \\
\end{tabular}
\end{center}
\end{table}

\section{Results and discussion}
\label{Sec: results_2}
\subsection{Influence of the aspect ratio}
\label{Sec: aspect ratio}

\begin{figure}[!ht]
\centering
\includegraphics[width=1.0\columnwidth]{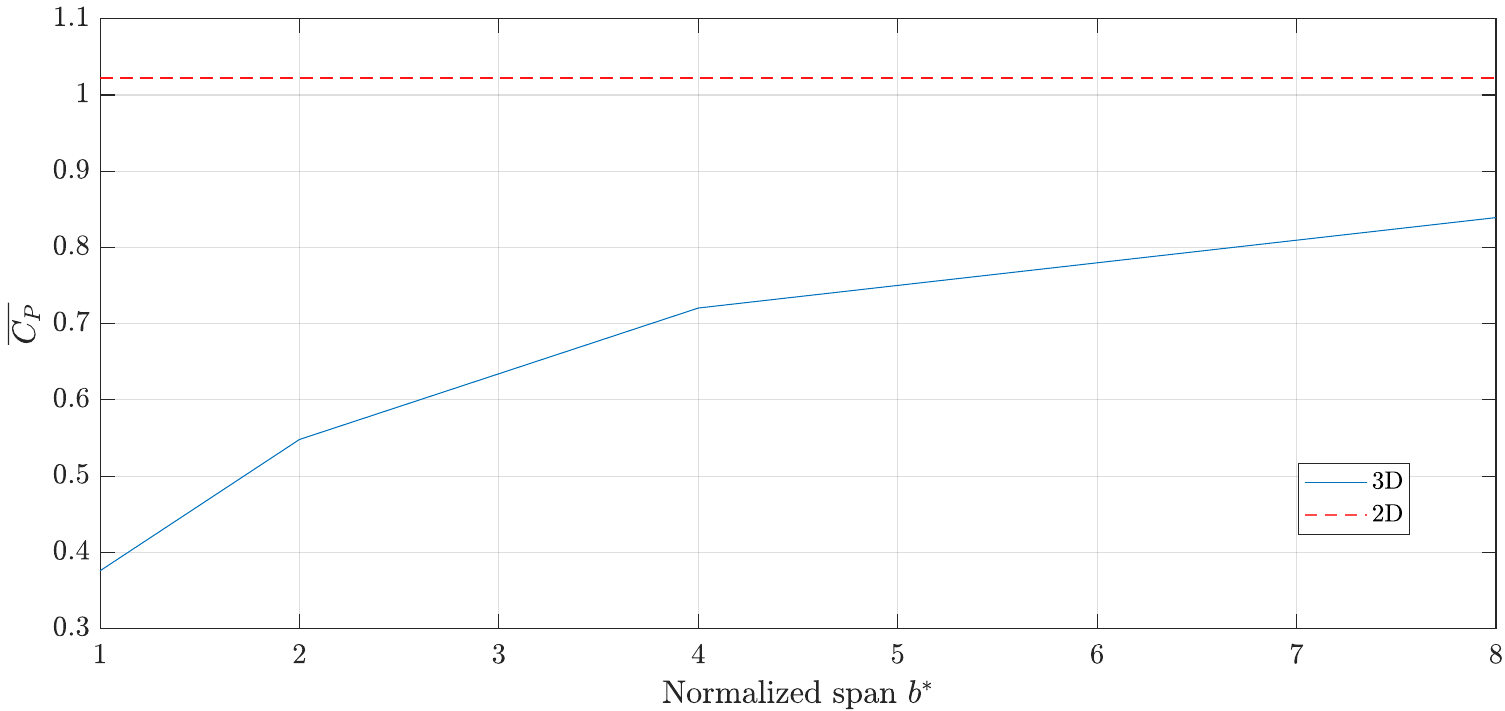}
\caption{Averaged power coefficient $\overline{C_P}$ as a function of the normalized span $b^*$.}
\label{fig_CP_3D}
\end{figure}

\begin{figure}[!ht]
\centering
\includegraphics[width=1.0\columnwidth]{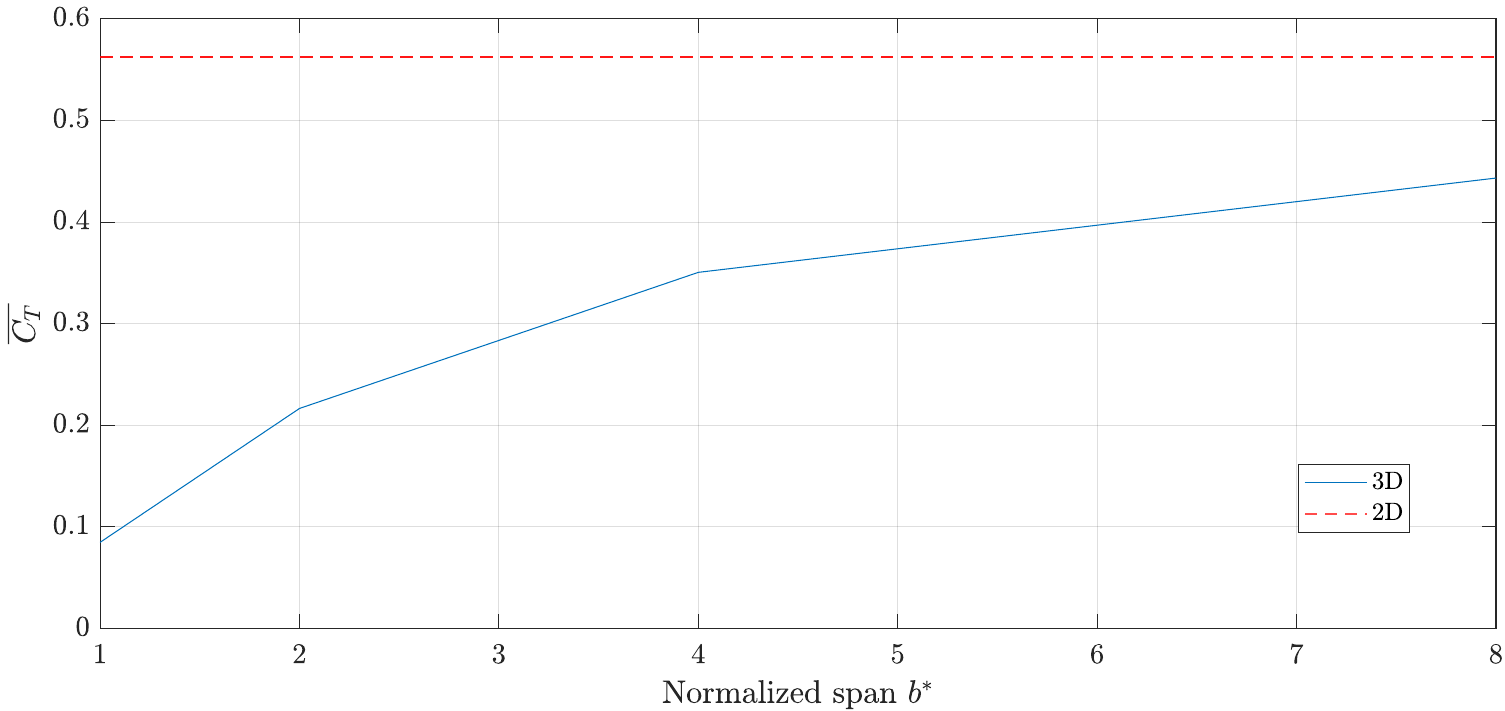}
\caption{Averaged thrust coefficient $\overline{C_T}$ as a function of the normalized span $b^*$.}
\label{fig_CT_3D}
\end{figure}

\begin{figure}[!ht]
\centering
\includegraphics[width=1.0\columnwidth]{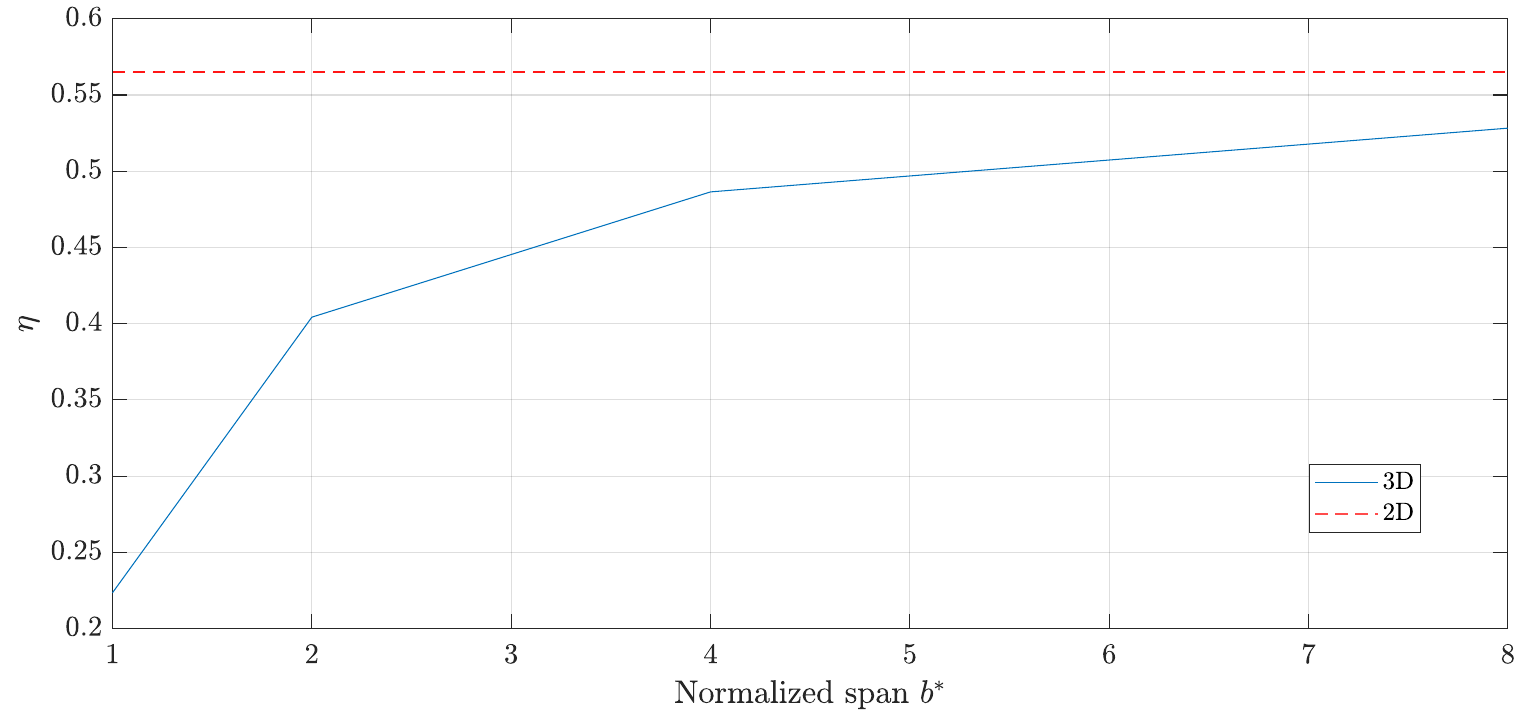}
\caption{Efficiency $\eta$ as a function of the normalized span $b^*$.}
\label{fig_eta_3D}
\end{figure}

This section investigates the effects of varying the aspect ratio of the wings.
This is performed by using the constant parameters defined in
\eq{eq_parametric_values_2} while modifying the normalized span $b^*$ between 1
and 8. The computed performances based on the metrics defined in \sect{Sec:
perf_metrics_2} are presented as a function of the normalized span and are
compared to the two-dimensional case, see \fig{fig_CP_3D} to \fig{fig_eta_3D}.
Cycle-averaged performance metrics are also summarized in
\tab{Tbl_performances}. \fig{fig_CP_3D} to \fig{fig_eta_3D} show that for wings
with an aspect ratio of 1, the cycle-averaged performances are severely
decreased compared to the ones of the two-dimensional case. As the normalized
span $b^*$ increases, these performances increase towards an asymptotic value
equivalent to the 2D results \citep{Papillon_clap_fling_2024}.  As the largest
normalized span considered in this study is 8, even better performances could
thus be obtained by considering wider wings.  \tab{Tbl_performances} also
summarizes these performance metrics, and it is seen that the relative
efficiency loss is indeed quite minimized by employing high aspect ratio wings
such as 4 and 8 ($\Delta \eta = -0.127$ and $\Delta \eta = -0.0534$,
respectively). Plates with a normalized span of 4 present a thrust coefficient
of $\overline{C_T}=0.351$ and an efficiency of 0.493. Plates with a higher
aspect ratio of 8 present a thrust coefficient of $\overline{C_T}=0.443$ and an
efficiency of 0.535. While these efficiencies are similar to the 2D result of
0.565, the thrust coefficient values of these two 3D cases are quite lower than
the one of $\overline{C_T}=0.563$.  The relative thrust loss for the two cases
$b^*=4$ and $b^*=8$ are -37.69\% and -21.17\%, respectively.
\begin{table}[htb]
\caption{Cycle-averaged performance metrics for 2D and 3D flapping wings.}
\begin{center}
\label{Tbl_performances}
\begin{tabular}{lccccc}
\toprule
 &
$b^*$ &
$\overline{C_T}$ & 
$\overline{C_P}$ & 
$\eta$ &  
$\Delta \eta$ relative to 2D (\%) \\
\midrule
3D & 1        & 0.0848 & 0.380 & 0.234 & -58.6 \\
3D & 2        & 0.217  & 0.555 & 0.402 & -28.8 \\
3D & 4        & 0.351  & 0.730 & 0.493 & -12.7 \\
3D & 8        & 0.443  & 0.851 & 0.535 & -5.34 \\
2D & $\infty$ & 0.563  & 1.022 & 0.565 & 0 \\
\bottomrule
\end{tabular}
\end{center}
\end{table}

\subsection{Three-dimensional effects and clap-and-fling }
\label{Sec: 3D effects}

\begin{figure}[!ht]
\centering
\includegraphics[width=1.0\columnwidth]{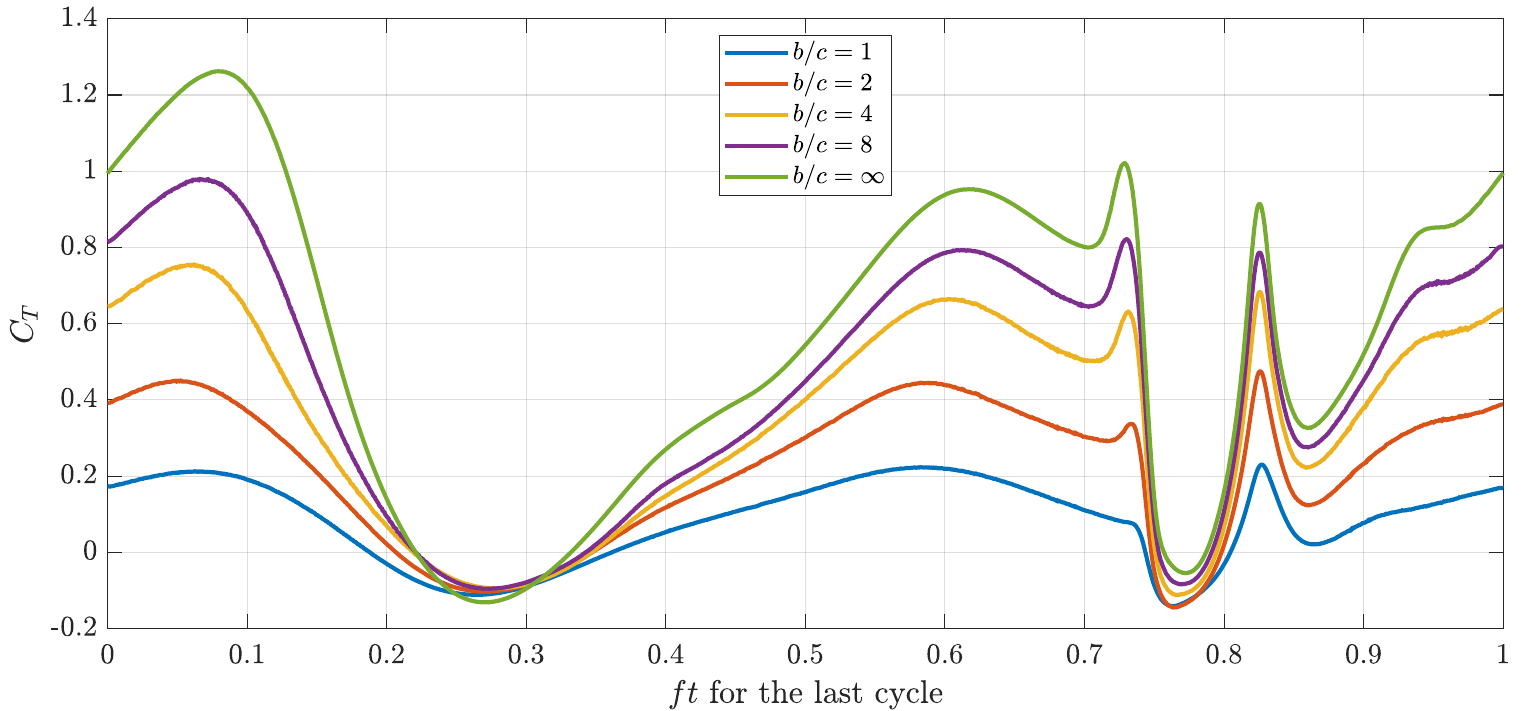}
\caption{Comparison of the instantaneous thrust coefficient $C_T$ for different normalized span $b^*$. }
\label{fig_inst_CT}
\end{figure}

\begin{figure}[!ht]
\centering
\includegraphics[width=1.0\columnwidth]{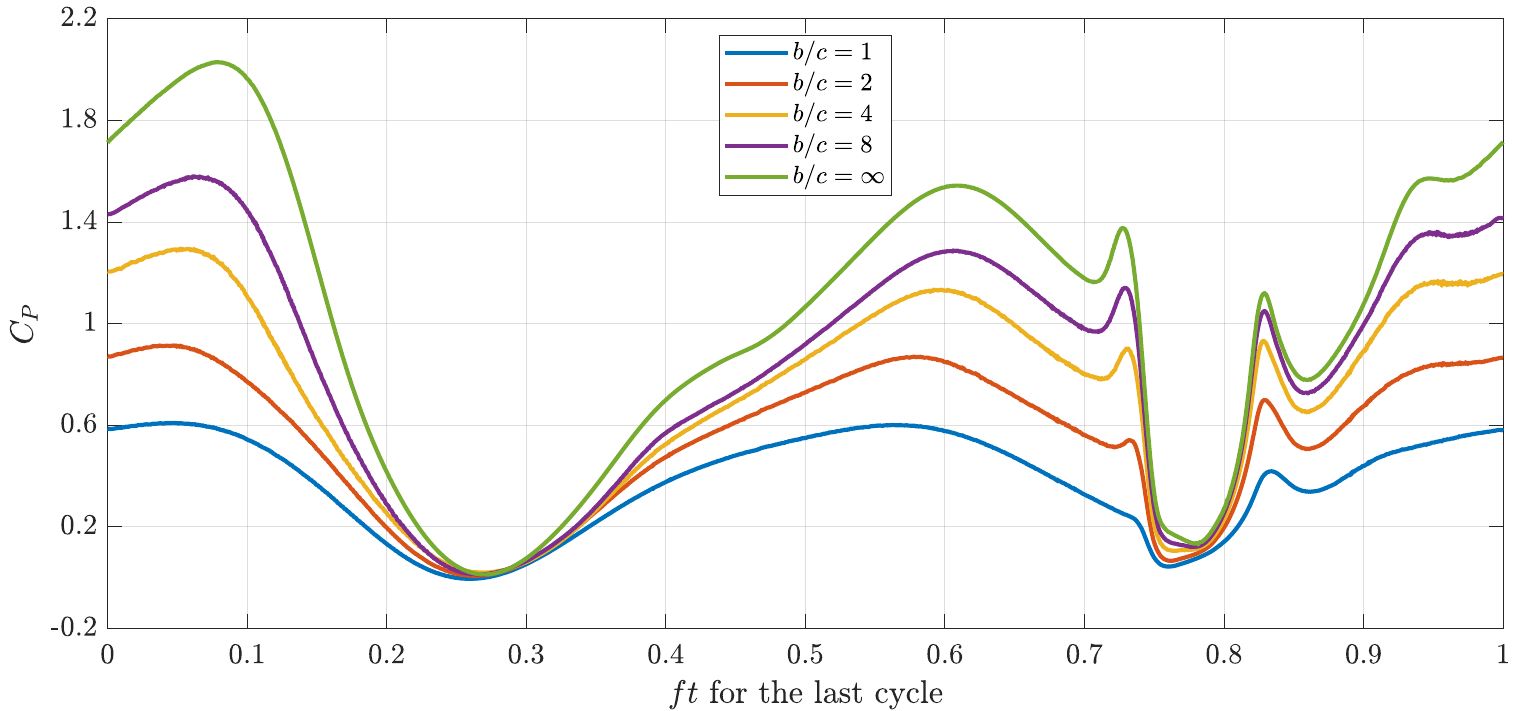}
\caption{Comparison of the instantaneous power coefficient $C_P$ for different normalized span $b^*$. }
\label{fig_inst_CP}
\end{figure}

Due to the importance of three-dimensional effects for flapping wings,
aerodynamic losses are expected relative to the reference two-dimensional case
computed by \cite{Papillon_clap_fling_2024}. To assess the instants in the cycle
these losses are most prevalent, the instantaneous thrust and power coefficients
are presented in \fig{fig_inst_CT} and \fig{fig_inst_CP}. As observed, the
three-dimensional cases tend to exhibit a global decrease in both thrust and
power coefficients relative to the two-dimensional case, as expected.  In this
analysis, two instants of interest will be discussed: $ft=0.600$ where the wings
are further apart with attached boundary layers, and $ft=0.826$ where the wings
reach the minimum spacing $d^*=0.005$ at their trailing edges with important
flow separation. The most noticeable impact is seen for this last instant where
a peak occurs due to the proximity of the wings. Decreasing the normalized span
$b^*$ greatly diminishes the thrust resulting from the strong interaction
between the wings, which is one of the main features of the clap-and-fling
phenomenon. As for the power coefficient, the peak occurring at the same instant
does not decrease as much when reducing the aspect ratio of the wings, which
implies that the instantaneous efficiency does not decrease linearly.

\begin{figure}[!ht]
\centering
\includegraphics[width=1.0\columnwidth]{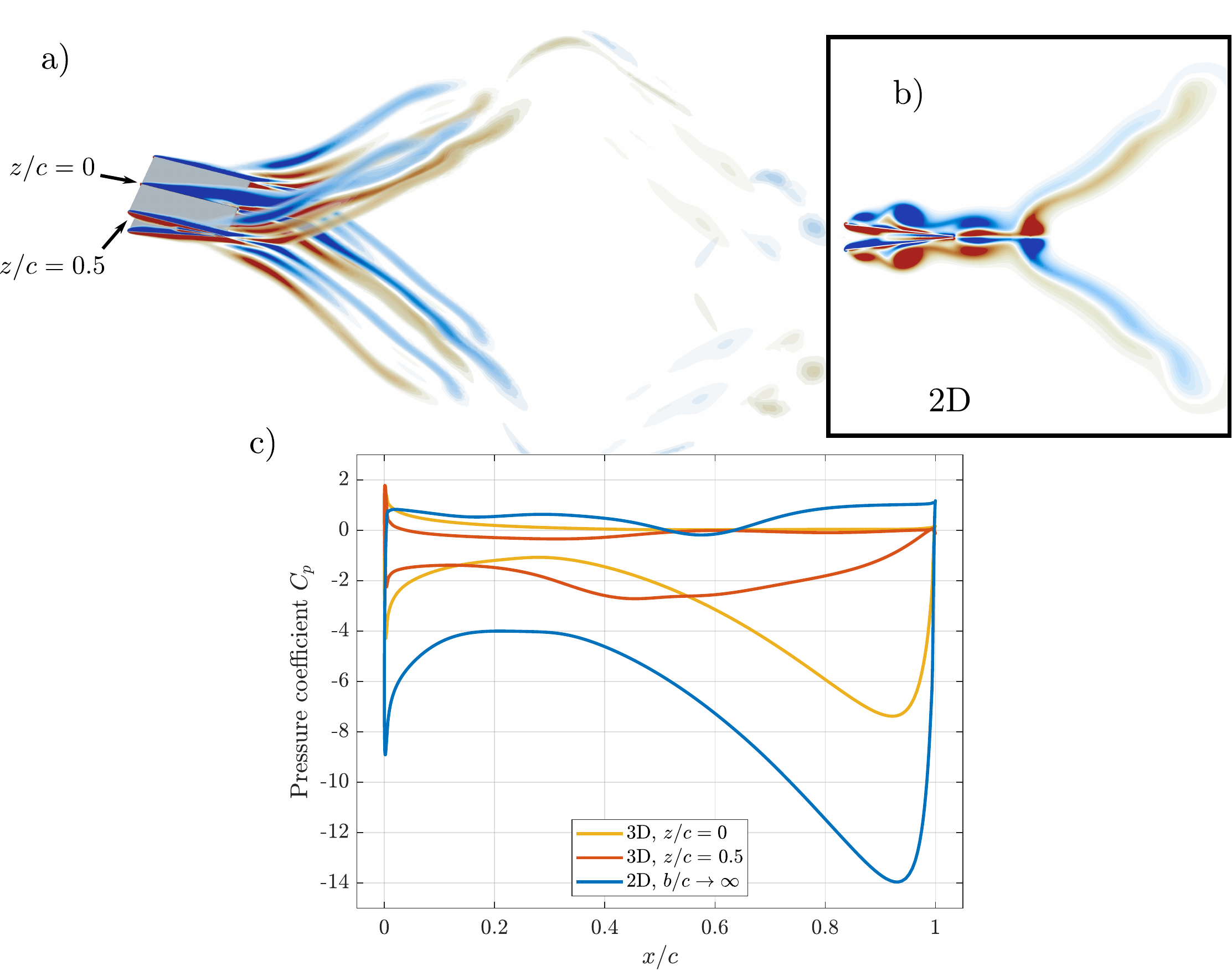}
\caption{Comparison of the normalized vorticity field ($\omega_z \, c/U_{\infty}$) and pressure coefficient distribution between 3D ($b^*=1$) and 2D at $ft=0.826$.}
\label{fig_vorticity_gap_1c}
\end{figure}

As mentioned, when the wings are close at their trailing edges, the
fluid between them is quickly ejected in the wake and carries two
trailing edge vortices. \cite{shyy_aerodynamics_2007} stated that this fast
ejection of fluid mitigates the Wagner effect, allowing each wing to
quickly build up circulation as they subsequently fling and thus increase the
produced thrust at this instant. This is well observed in
\fig{fig_vorticity_gap_1c}(b) for the two-dimensional case where the
trailing-edges vortices are indeed being quickly ejected. Additionally, while
the wings fling, fluid is sucked into the gap as they separate which results
in a fast buildup of circulation of opposite signs on each wing. This results
in the formation of strong leading-edge vortices along with flow separation in
the boundary layers between the wings. A low-pressure zone is then generated
at the core of these vortices, which greatly increases the thrust due to the
favorable inclination of the wings. As observed in
\fig{fig_vorticity_gap_1c}(a) and \fig{fig_vorticity_gap_1c}(b), the
vortices present in the two-dimensional case are essentially non-existent in
the 3D case with $b^*=1$ at the instant $ft=0.826$. As the wings then
separate, the leading-edge vortices subsequently forming between the wings are
also absent, which supports the fact that the thrust observed in
\fig{fig_inst_CT} is much lower.

When the wings are inclined to obtain the minimum spacing $d^*$ at their
trailing edges, an important low-pressure zone is present between them.
This is observed in the 2D curve from \fig{fig_vorticity_gap_1c}(c) which
presents comparisons of the instantaneous pressure coefficient distribution of
the upper wing between 2D and 3D at two different positions $z/c$ along the
span. Indeed, the inner surface of the upper wing presents an important
low-pressure coefficient $C_p$ at this instant, while the outer surface has a
low positive value which does not fluctuate much. For the three-dimensional
case, this difference in pressure between the inner and outer surfaces must
equilibrate at each wing-tip, which produces wing-tip vortices through the
generated secondary flow. These wing-tip vortices are oriented such
that an upwash occurs along the span of the upper wing (the opposite occurs
for the lower wing). In the two-dimensional case, the magnitude of the
effective angle of attack of the wings resulting from the combined heaving,
pitching, and deviation motion is increased when the minimum spacing is
reached, which enhances the thrust \citep{Papillon_clap_fling_2024}.
However, in the three-dimensional cases, the
upwash modifies the effective vertical velocity, which greatly decreases the
magnitude of the effective angle of attack as the wing span is progressively
reduced. As a result, this alteration of the effective angle of attack, which
is closer to $0 \degree$, generates an induced drag in the opposite direction of
the thrust, reducing substantially the latter. The thrust and
power coefficient are thus greatly affected as seen in \fig{fig_inst_CT} and
\fig{fig_inst_CP}. 

This alteration to the effective angle of attack when the wings are in proximity
at $ft=0.826$ also affects the pressure distribution as seen in
\fig{fig_vorticity_gap_1c}(c) and \fig{fig_vorticity_gap_1c}(d). At $z/c=0$,
although the 2D and 3D curves of the outer surface of the upper wing do not
follow the same evolution, they exhibit similar values. However, the pressure
distribution of the inner surface, while following a similar trend, undergoes a
positive shift for the 3D case relative to the 2D one. At the wing tips
($z/c=0.5$), the two pressure distributions are completely different, as seen in
\fig{fig_vorticity_gap_1c}(c), due to presence of wing-tip vortices which are
very close. This important variation of the pressure distribution along the span
of the wings explains why well correlated spanwise vortices such as the ones
observed in \fig{fig_vorticity_gap_1c}(b) are not seen in the 3D simulation.

\begin{figure}[!ht]
\centering
\includegraphics[width=1.0\columnwidth]{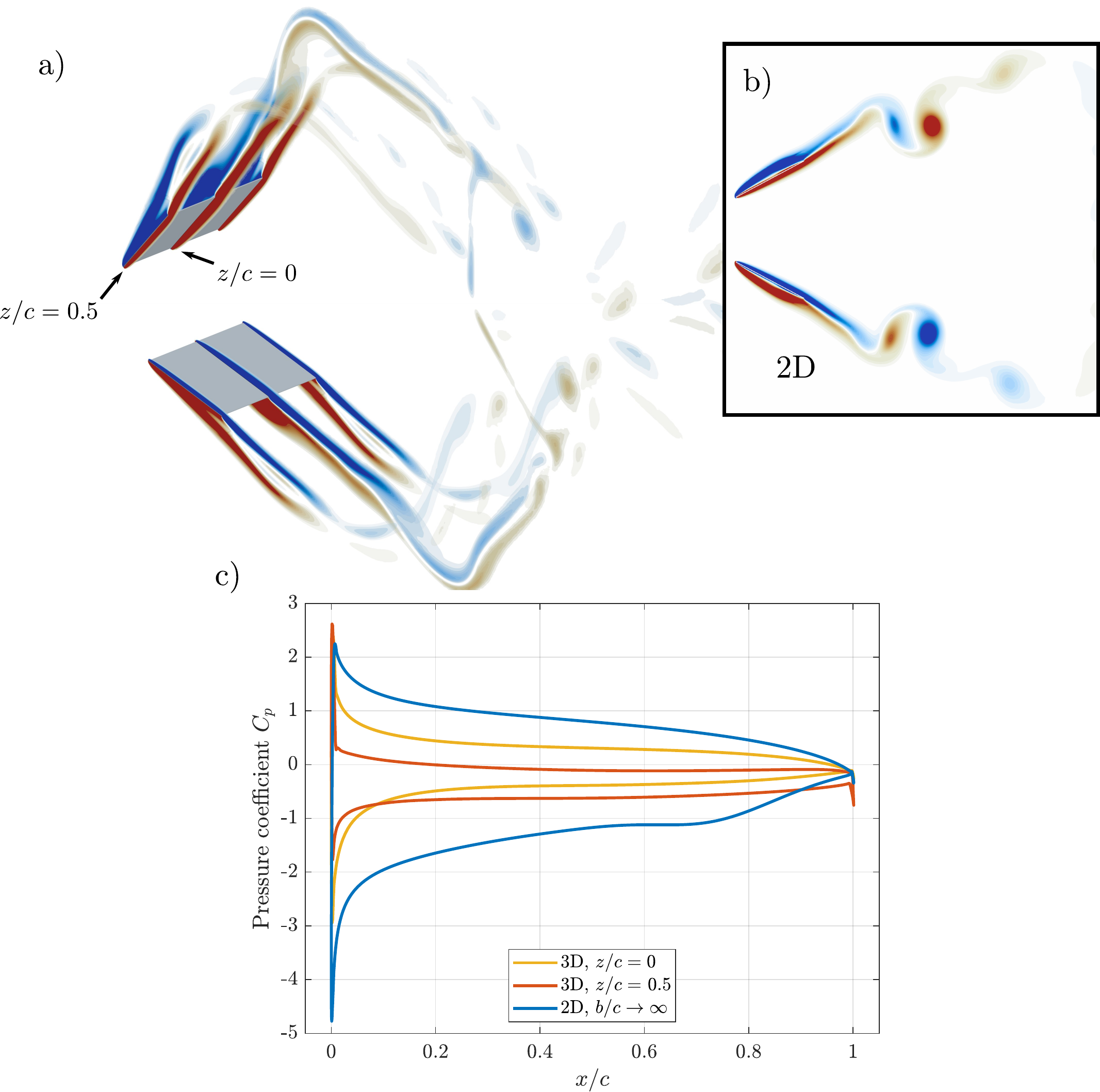}
\caption{Comparison of the normalized vorticity field ($\omega_z \, c/U_{\infty}$) and pressure coefficient distribution between 3D ($b^*=1$) and 2D at $ft=0.600$.}
\label{fig_vorticity_attached_1c}
\end{figure}

\fig{fig_vorticity_attached_1c} shows similar normalized vorticity fields and
pressure distributions of the wings during the upstroke at $ft=0.600$ for a
case where the boundary layers are attached when the wings are further apart.
As observed, the normalized vorticity fields of this case are much more similar
even though significant three-dimensional effects occur due to the finite span
of the 3D wings considered. Therefore, while the pressure coefficient
distributions of the 3D case follow a similar distribution to the
two-dimensional simulation, it is much closer to zero in general. Additionally,
the pressure distribution varies less along the spanwise direction of the 3D
case, which can be attributed to the absence of important flow separation.
Nonetheless, 3D effects are still important and diminish the spread of the
pressure distribution, which necessarily affects the aerodynamic performances
observed at this instant. Indeed, the instantaneous thrust coefficient $C_T$
and power coefficient $C_P$ are much lower for $b^*=1$ at $ft=0.600$ relative
to the two-dimensional case. 

\begin{figure}[!ht]
\centering
\includegraphics[width=1.0\columnwidth]{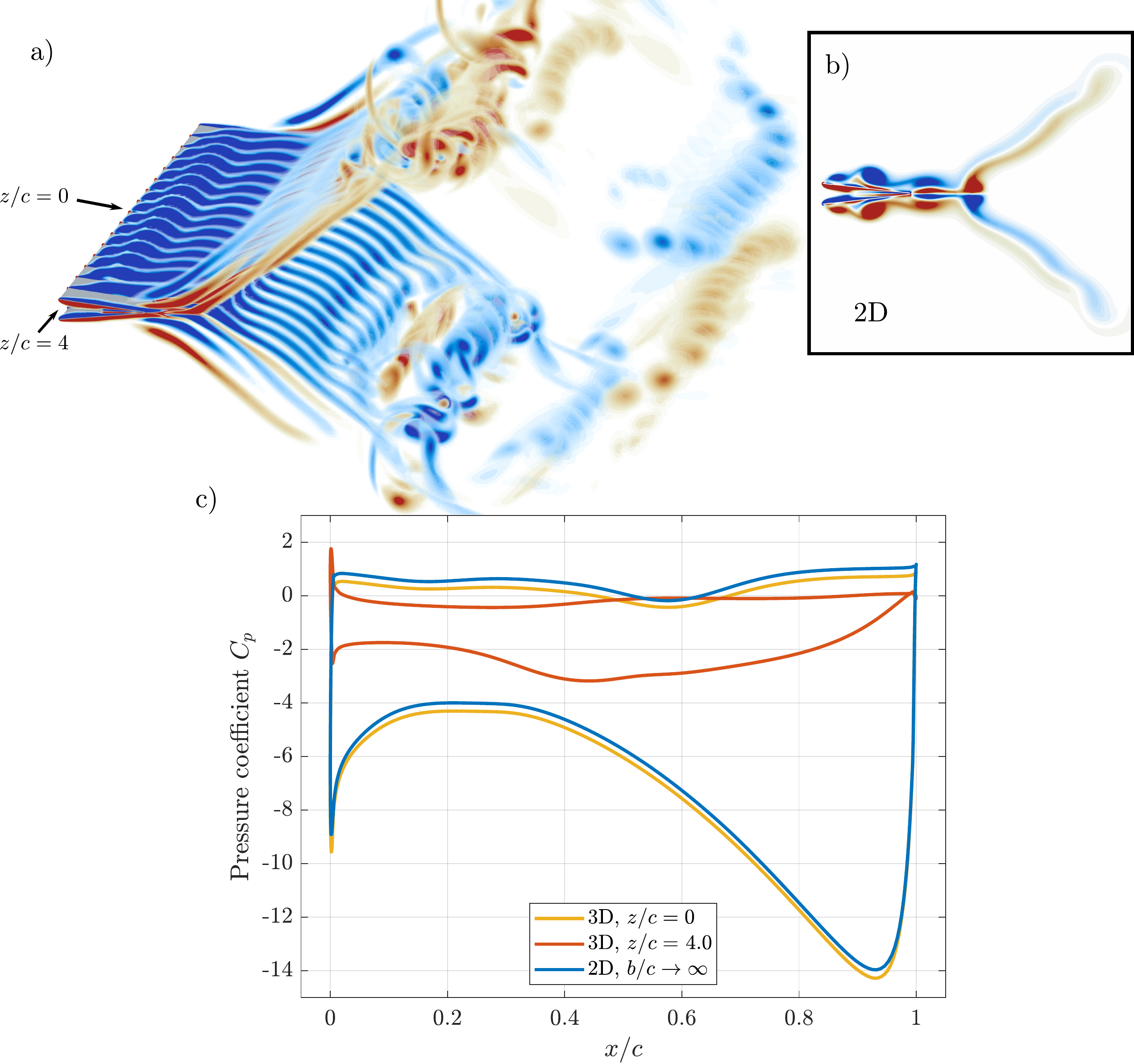}
\caption{Comparison of the normalized vorticity field ($\omega_z \, c/U_{\infty}$) and pressure coefficient distribution between 3D ($b^*=8$) and 2D at $ft=0.826$.}
\label{fig_vorticity_gap_8c}
\end{figure}

Next, \fig{fig_vorticity_gap_8c} presents a comparison of the normalized
vorticity fields and pressure coefficient distributions for 3D wings with
$b^*=8$ for the same instant $ft=0.826$ where the minimum spacing $d^*$ is
reached at the trailing edges. The vorticity distribution shown in the figure
at several spanwise positions is much more similar to the equivalent
two-dimensional case, especially at the midspan. Indeed, flow separation does
occur in the same fashion at $z/c=0$ with the development of similar
leading-edge vortices, which were absent in the previous results with a span
$b^*=1$. Moreover, the pressure coefficient distribution at the midspan is
almost identical to the one of the two-dimensional case, as seen on
\fig{fig_vorticity_gap_8c}(c), which confirms that flow is less affected by
the upwash previously mentioned with the increase in span. This can be
explained by the proximity of wing-tip vortices of opposite sign from
each wing which counteract each other at this instant, thus resulting in an
importantly diminished upwash. As for the one at the wing tip ($z/c=2.0$), the
pressure distribution remains similar to the one produced by the smaller wing and
differs from the 2D result, which is expected since three-dimensional effects
originate from wing tips. Nonetheless, it is observed that proper kinematics
combined with the wing proximity inherent in the clap-and-fling mechanism
attenuate three-dimensional effects during the clap phase, which is beneficial
for thrust generation and efficiency. This is even more prevalent for cases
where the foils' span is sufficiently increased as 3D effects are even less
important.

\begin{figure}[ht!]
\centering
\includegraphics[width=1.0\columnwidth]{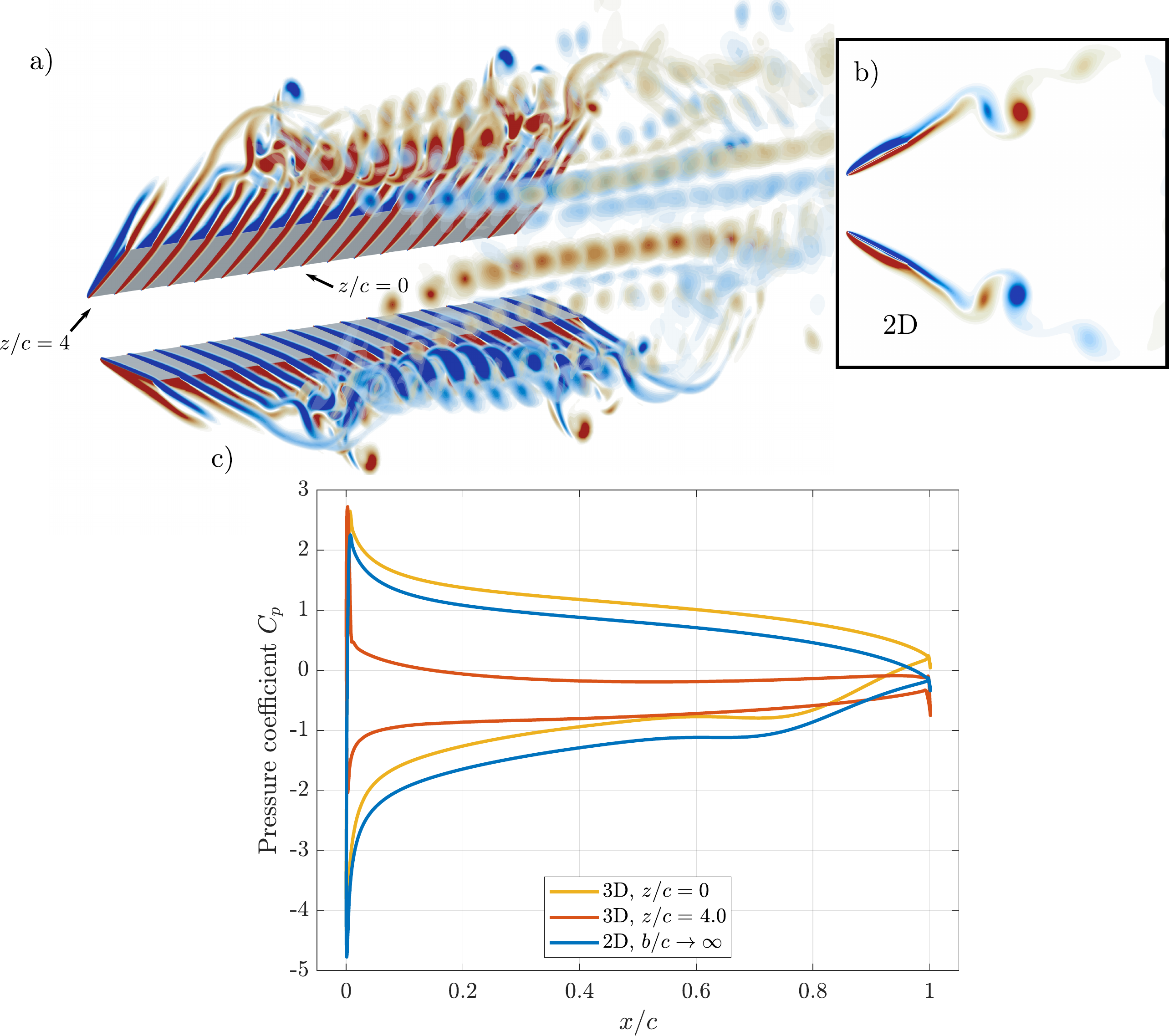}
\caption{Comparison of the normalized vorticity field ($\omega_z \, c/U_{\infty}$) and pressure coefficient distribution between 3D ($b^*=8$) and 2D at $ft=0.600$.}
\label{fig_vorticity_attached_8c}
\end{figure}

Lastly, \fig{fig_vorticity_attached_8c} presents results for wings with a span
of $b^*=8$ at $ft=0.600$. At this moment, this configuration presents attached
boundary layers. It is also shown that when the wings are close ($ft=0.826$),
using wings with a normalized span as low as $b^*=1$ yielded results much more
similar to the 2D prediction.  Indeed, while the normalized vorticity fields are
quite similar to the 2D results at $ft=0.600$, they are even more so in this
case presented in \fig{fig_vorticity_attached_8c}(a). Boundary layers remain
attached as in the two-dimensional case, especially at the midspan $z/c=0$.
However, noticeable differences in the pressure coefficient distributions
presented are observed in \fig{fig_vorticity_attached_8c}(c).  At the wings'
tips $z/c=4.0$, 3D effects are significant, resulting in a near-zero pressure
difference between the upper and lower sides of the wing.  These effects also
reduce the spread of thrust distribution along the span (discussed in \sect{Sec:
Distribution}).  As for the midspan $z/c=0$, the chordwise distribution is much
more similar to the two-dimensional simulation. However, a small positive shift
is observed for both curves related to the inner and outer surface of the wing
due to 3D effects. Indeed, a downwash (for the upper wing) is generated by the
wing tip vortices at this instant, which especially affects the outer surfaces
through the direction of the induced velocity.  This behavior is expected since
the wings are further apart at this instant.  Therefore, the downwash and upwash
velocity induced by each wing tip's vortices do not cancel each other out in the
same way as when the wings are closer during the clap phase.

In summary, three-dimensional effects can be attenuated during the clap phase
of the motion due to the proximity of each wing tip vortex, resulting in the
cancellation of their respective induced velocities, which is beneficial for
thrust generation and efficiency. This is even more prevalent in cases where
the wing span is sufficiently large as 3D effects are then less important.

%

\subsection{Thrust and power spanwise distribution}
\label{Sec: Distribution}
\begin{figure}[!ht]
\centering
\includegraphics[width=1.0\columnwidth]{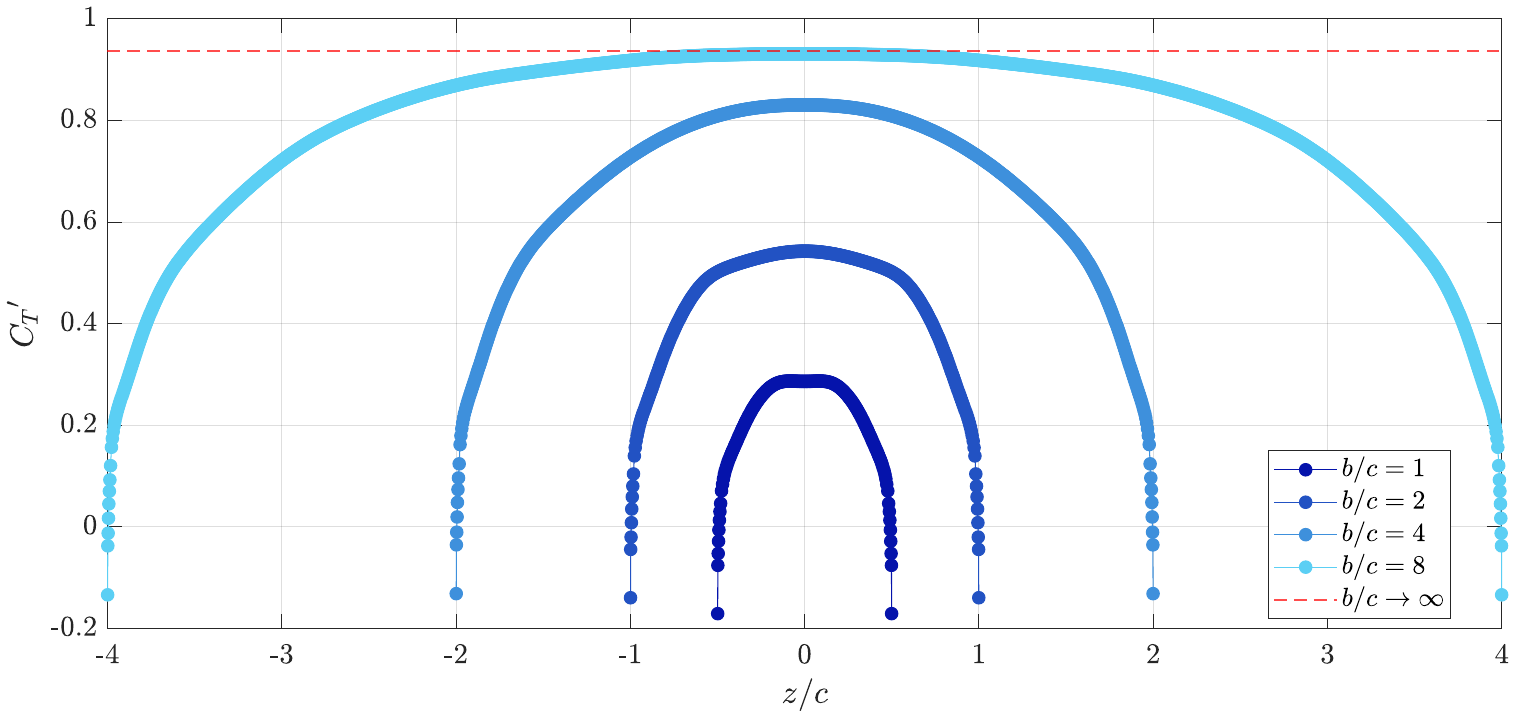}
\caption{Comparison of the propulsive spanwise distribution ${C_T}'$ for different normalized span $b^*$ at $ft=0.600$.}
\label{fig_CT_3D_loading_attached}
\end{figure}

\begin{figure}[!ht]
\centering
\includegraphics[width=1.0\columnwidth]{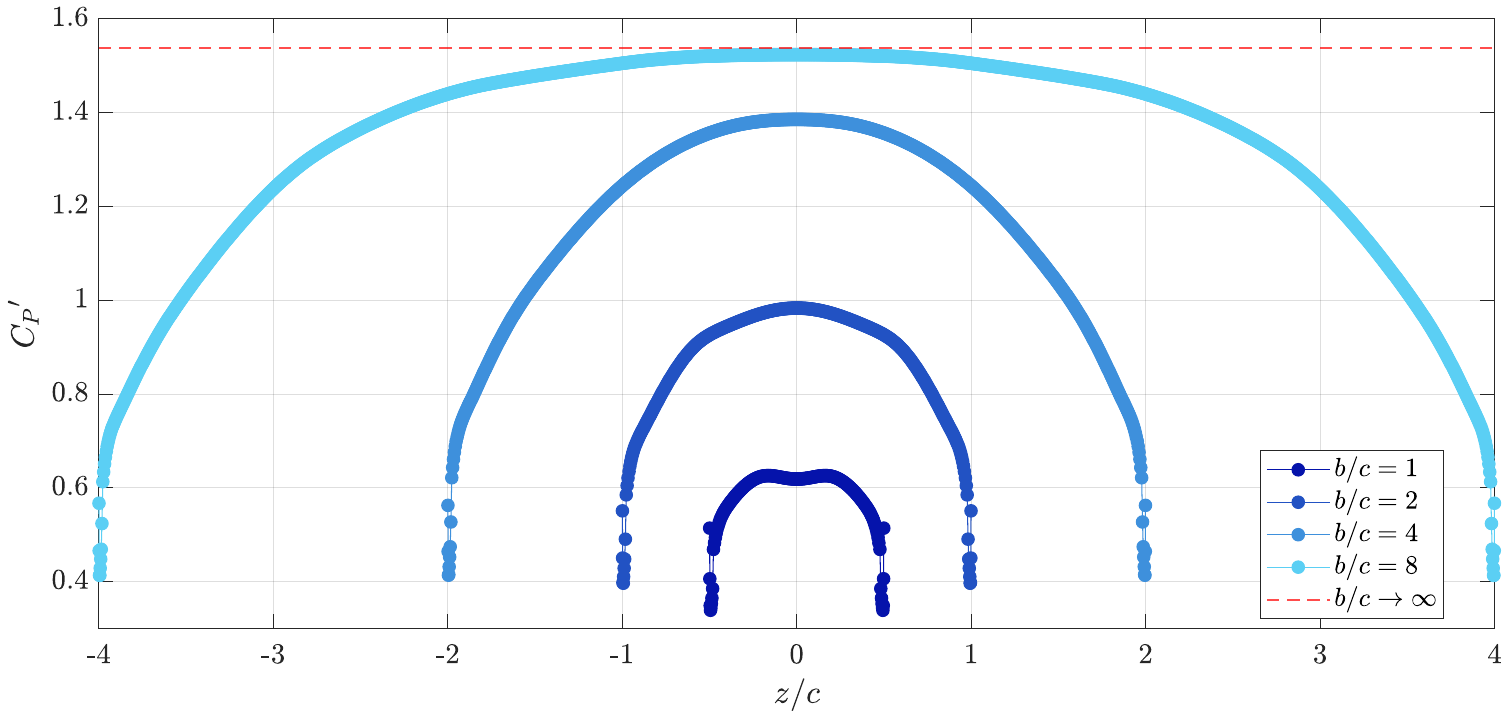}
\caption{Comparison of the power spanwise distribution ${C_P}'$ for different normalized span $b^*$ at $ft=0.600$.}
\label{fig_CP_3D_loading_attached}
\end{figure}

As mentioned in \sect{Sec: 3D effects}, the clap-and-fling mechanism attenuates
adverse 3D effects, which increases the thrust and propulsive efficiency as the
normalized span of the wings increases. Thus, this section aims to further
analyze the impact of these three-dimensional effects, specifically on the
wing loading distribution. \fig{fig_CT_3D_loading_attached} and
\fig{fig_CP_3D_loading_attached} present the propulsive and power spanwise
distributions for several normalized wing spans at $ft=0.600$. This corresponds
to an instant when the wings are far from each other and their boundary layers
remain attached. As seen in those two figures, three-dimensional effects
are most important near the end of the wings where wingtip vortices are at
their strongest. This diminishes significantly the local propulsive thrust
coefficient ${C_T}'$, which becomes negative. As for the local power thrust
coefficient ${C_P}'$, its value also decreases notably towards zero. It
is known that the lift and aerodynamic moment $M_z$ at the wing tips are null
because the pressure difference between the intrados and the extrados cannot be
maintained, which explains the distribution obtained in
\fig{fig_CP_3D_loading_attached}. Additionally, the performance metrics are
maximum at the middle of the wings and increase with the normalized span $b^*$.
Indeed, as the wing aspect ratio increases towards higher values, the values of
${C_T}'$ and ${C_P}'$ at the wing midspan reach a value essentially
identical to the equivalent two-dimensional case, which is expected. However, a
phenomenon occurs for wings with lower aspect ratios where 3D effects are more
important. As the local thrust and power coefficients increase towards the
middle of the wings, an abrupt plateau occurs, which reduces their maximum
magnitude. This is especially noticeable for the case with the lowest
normalized span $b^*=1$ in \fig{fig_CP_3D_loading_attached}. 

\begin{figure}[!ht]
\centering
\includegraphics[width=\columnwidth]{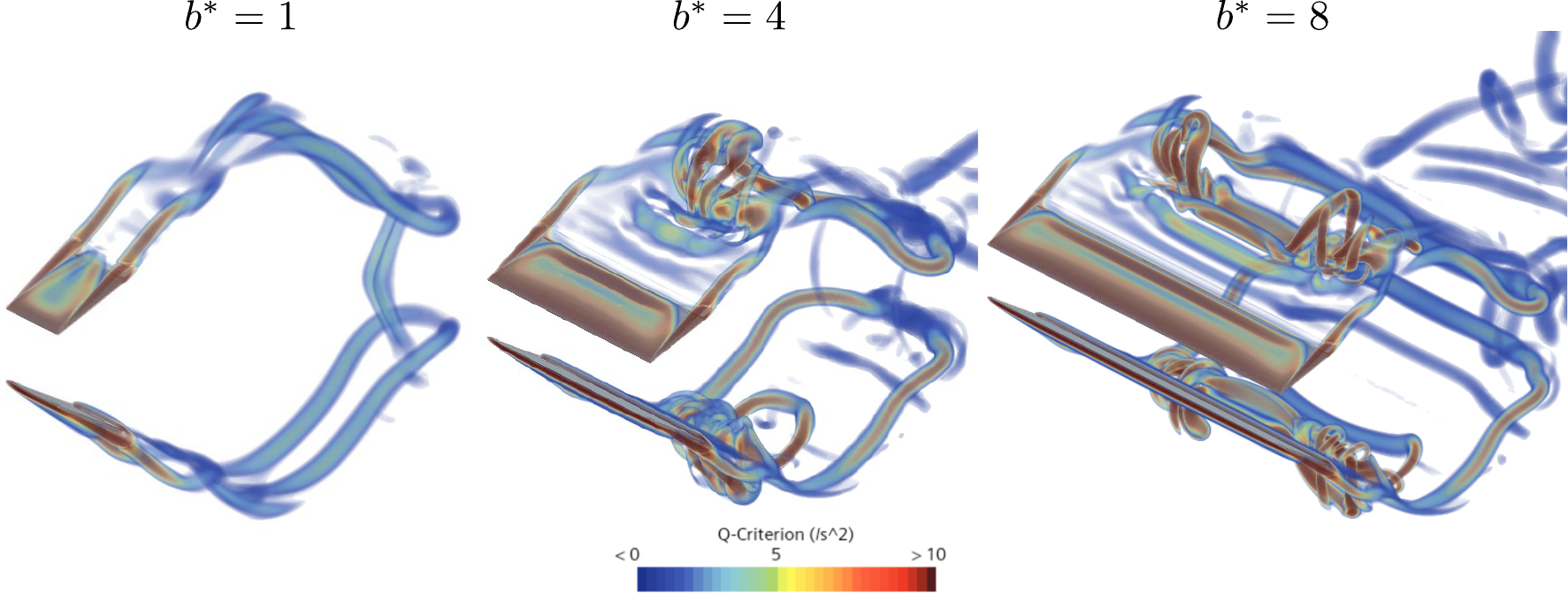}
\caption{Comparison of the Q-criterion field between the cases $b^*=1$, $b^*=4$ and $b^*=8$ at $ft=0.600$.}
\label{fig_Q_criterion_attached}
\end{figure}

\fig{fig_Q_criterion_attached} presents a comparison of the Q-criterion field
for the cases $b^*=1$, $b^*=4$ and $b^*=8$, which allows clear visualization of
vortical structures around the wings. In each case, tip vortices are noticeable
as expected.  For higher aspect-ratio wings, significant vortices are present
near their trailing edges within the boundary layers, as indicated by the
Q-criterion contours.  Those swirling structures with a predominant vorticity
component oriented in the spanwise direction are well correlated along the span.
Consequently, the boundary layers remain attached, which leads to the lift and
power coefficient distributions observed in \fig{fig_CT_3D_loading_attached} and
\fig{fig_CP_3D_loading_attached}.  Consequently, these distributions tend toward
the two-dimensional results near the midsection of the wings. 

In \fig{fig_Q_criterion_attached}, the boundary layers roll up into spanwise
vortices near the external surfaces of the wings due to the existing pressure
gradient which causes flow separation in the same manner as the equivalent 2D
case. These vortices are then ejected in the wake, and three-dimensional
instabilities develop due to the proximity of wing-tip vortices. For the $b^*=4$
and $b^*=8$ cases, these instabilities eventually cause a breakdown of the
vortices which were initially well correlated in the spanwise direction,
producing complex vortical structures observed in
\fig{fig_Q_criterion_attached}. For the wings with $b^*=4$, the span is still
not large enough to yield a vorticity distribution similar to the case $b^*=8$.
For the former, the vortices present in the wake are regrouped near the midspan
since the wings are not long enough in the spanwise direction. For the case with
the largest span $b^*=8$, the vortical structures are distributed symmetrically
around the midspan but do not intersect with each other. As for the wings with
the normalized span $b^*=1$, the vortex distribution is quite different. While
similar wing-tip vortices are present, their intensity which is proportional to
the generated lift, is quite reduced for such a compact wing. This is confirmed
by the power coefficient distribution presented in
\fig{fig_CP_3D_loading_attached} which strongly depends on the lift of the wings
and is, accordingly, much lower than the other curves.  Moreover, the spanwise
distribution of vortical structures is not as correlated as in the two other
cases. Indeed, for such low aspect ratio wings, three-dimensional effects are
more important due to the proximity of the wing-tip vortices which reduce the
intensity of the generated spanwise vortices. These vortices resulting from the
roll-up of shear layers rapidly lose their spanwise correlation due to the
strong wing-tip vortices oriented in the chordwise direction, as seen in
\fig{fig_Q_criterion_attached}. Therefore, the presence of swirling structures
in the wake shed from the previous motion cycle is much less important for the
lower aspect ratio case, which significantly changes the behavior of the flow.
As a result, the thrust and power spanwise distribution are affected as
mentioned above with a plateau occurring near the midsection for such an instant
where the wings are far from each other ($ft=0.600$).

\begin{figure}[!ht]
\centering
\includegraphics[width=1.0\columnwidth]{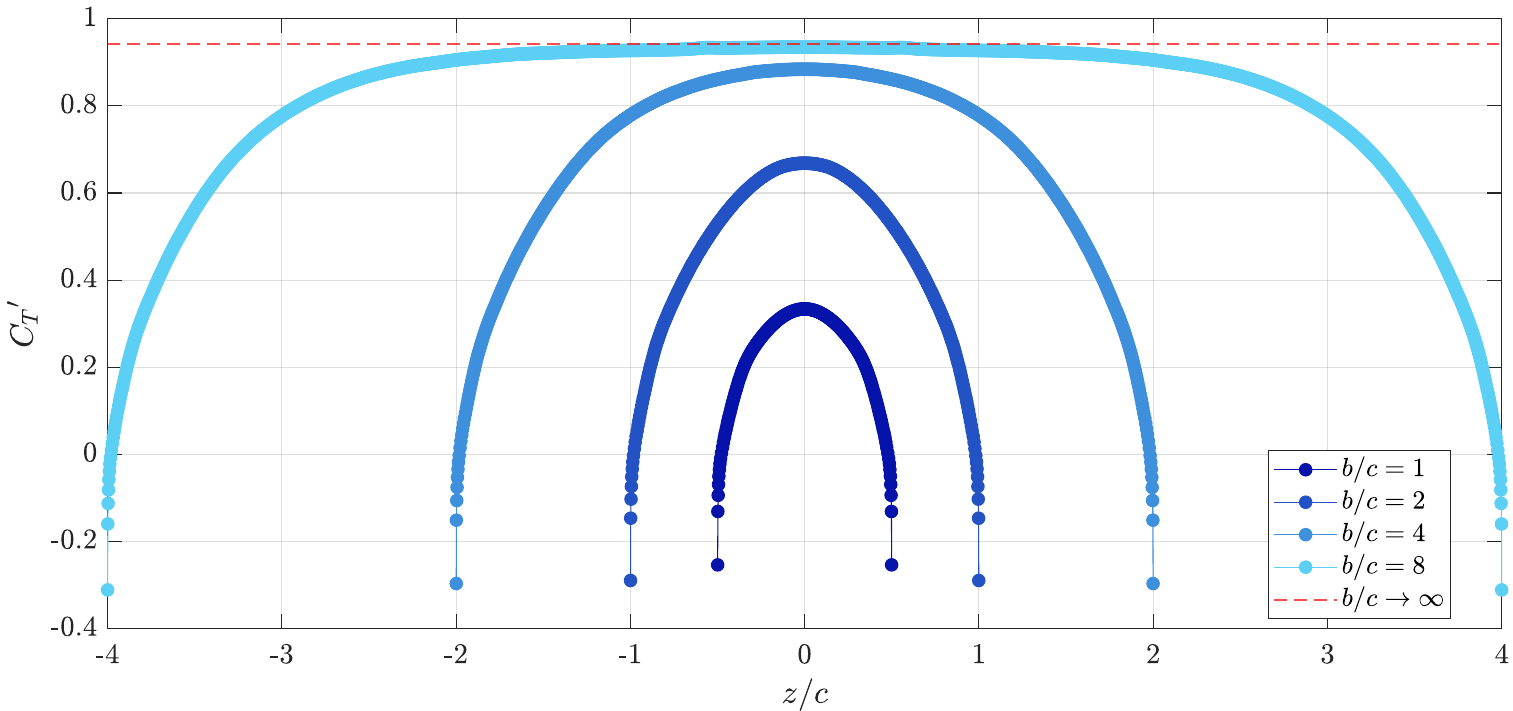} 
\caption{Comparison of the propulsive spanwise distribution ${C_T}'$ for different normalized span $b^*$ at $ft=0.826$.}
\label{fig_CT_3D_loading_detached}
\end{figure}

\begin{figure}[!ht]
\centering
\includegraphics[width=1.0\columnwidth]{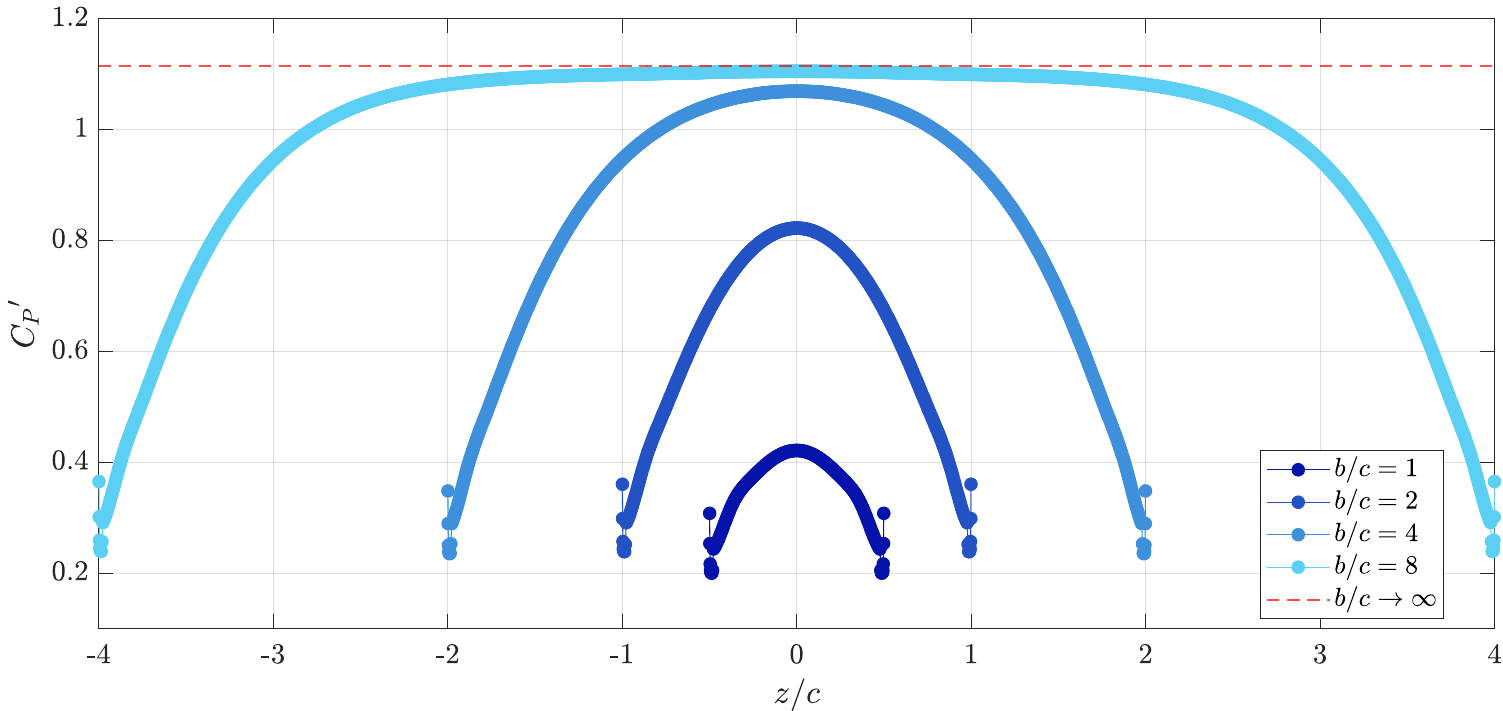} 
\caption{Comparison of the power spanwise distribution ${C_P}'$ for different normalized span $b^*$ at $ft=0.826$.}
\label{fig_CP_3D_loading_detached}
\end{figure}

\begin{figure}[!ht]
\centering
\includegraphics[width=1.0\columnwidth]{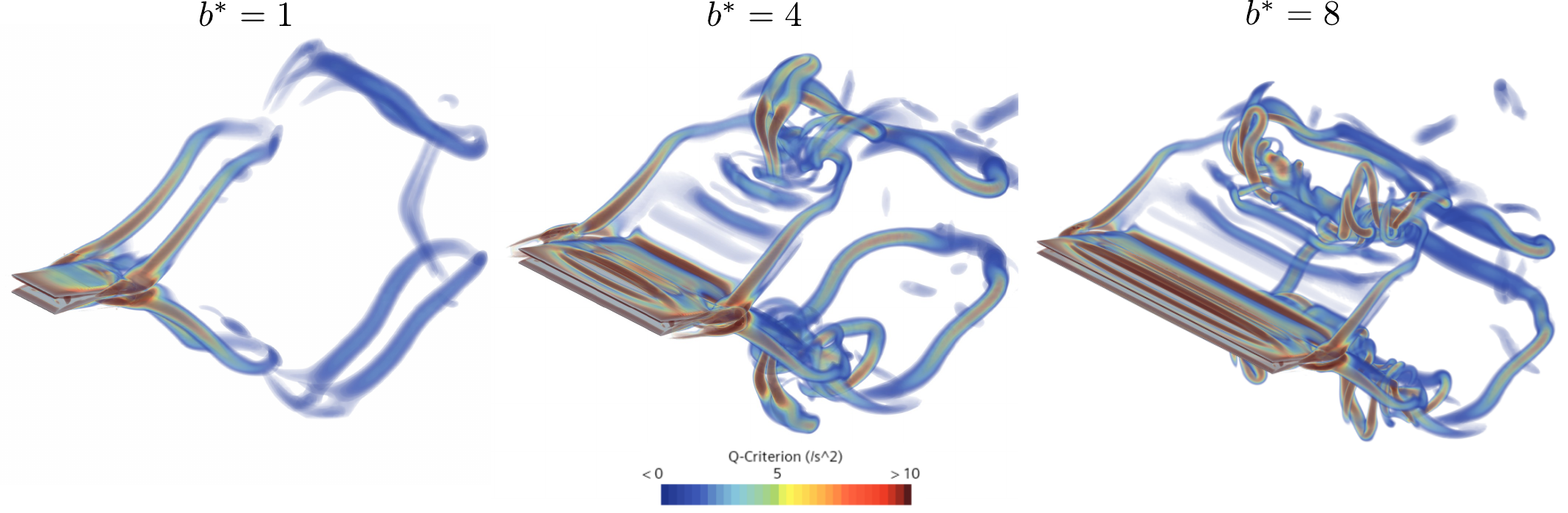}
\caption{Comparison of the Q-criterion field between the cases $b^*=1$, $b^*=4$ and $b^*=8$ at $ft=0.826$.}
\label{fig_Q_criterion_detached}
\end{figure}

The propulsive and power spanwise distributions for the second instant of
interest ($ft=0.826$) are investigated. At this instant, the wings are in
proximity with the trailing edges reaching the minimum spacing $d^*=0.005$,
resulting in a local peak for both the thrust and power coefficients.  Similar
observations as the previous ones at $ft=0.600$ can be made concerning
\fig{fig_CT_3D_loading_detached} and \fig{fig_CP_3D_loading_detached}. Indeed,
the local ${C_T}'$ and ${C_P}'$ diminishes greatly near the wing tips where 3D
effects are dominant. Again, the performance metrics are maximum at the middle
of the wings and increase towards the two-dimensional equivalent value with the
normalized span $b^*$. It is observed that for this case with high proximity of
the foils, those thrust and power coefficients peaks increase more quickly
towards the asymptote for all aspect ratios considered in contrast to results
from \fig{fig_CT_3D_loading_attached} and \fig{fig_CP_3D_loading_attached}. This
can be explained by the proximity of the foils which mitigates three-dimensional
effects, especially towards the midsection of the wings, as explained in
\sect{Sec: 3D effects}.  \fig{fig_Q_criterion_detached} presents the vortical
structures of the present case for the normalized span $b^*=1$, $b^*=4$, and
$b^*=8$. As observed, the proximity of the wings implies an important
confinement of the fluid which begins to accelerate between the wings as they
fling apart and leads to the ejection of trailing edge vortices in the wake.
This was observed in the 2D results by \cite{Papillon_clap_fling_2024} where a
confinement effect was generated and greatly increased the produced thrust
through the resultant low-pressure zone present between the wings.

Furthermore, the confinement effect is known to limit the vertical component of
the airflow around the wing tips and disrupt the generated wing-tip vortices.
Hence, the downwash intensity is reduced and the aerodynamic performances are
increased, which corroborates the results of \cite{blockley_ground_2010}. In the
present case, while the wake structure of the three cases remains quite
different, the proximity of the two vortices of opposite signs near each wingtip
counteracts each other during the clap phase of the motion in the manner of a
ground effect. In this configuration, this results in an importantly diminished
upwash in the boundary layers between the wings, which limits the aerodynamic
performance losses. Indeed, the spanwise force distribution is not as affected
by the three-dimensional geometry of the wings, which explains why no plateau in
the propulsive and power spanwise distribution is observed in
\fig{fig_CT_3D_loading_detached} and \fig{fig_CP_3D_loading_detached}. In short,
the clap-and-fling mechanism attenuates the three-dimensional effects and
increases thrust performances, as shown by the previous spanwise distributions,
especially when the wings are near each other during the motion.

\section{Conclusion}
\label{Sec: conclusion}
This study investigated oscillating wings utilizing the clap-and-fling mechanism
for thrust generation. Three-dimensional numerical simulations were carried out
at a Reynolds number of 800, employing plate-shaped wings with sinusoidal
pitching, heaving, and deviation motions. Specific phase shifts between these
motions and other kinematic parameters were prescribed in the numerical
experiments according to results obtained from two-dimensional studies. The
study aimed to evaluate the influence of three-dimensional effects on the
system's performance by employing wings with different aspect ratios.  

By considering wings with a normalized span as high as 8, thrust and efficiency
performance reductions were mitigated.  Plates with an aspect ratio of 4 present
a thrust coefficient of $\overline{C_T}=0.351$ and an efficiency of 0.493. As
for the highest aspect ratio considered, which is of 8, a thrust coefficient of
$\overline{C_T}=0.443$ and an efficiency of 0.535 were reached. The latter case
represents losses for the thrust coefficient of -21.17\% and -5.35\% for the
efficiency relative to the 2D results.  Accordingly, it can be asserted that
with a significantly high aspect ratio, the efficiency losses can be much
reduced, but the thrust production of the system remains greatly affected by the
three-dimensional effects.

Additionally, it was shown that the adverse 3D effects were more preponderant
when the wings were further apart during the motion. Near this instant
($ft=0.600$), a downwash relative to the upper wing is generated by the wing tip
vortices on its external surface due to the pressure difference between its
extrados and intrados. This greatly affects the effective angle of attack of the
wings and thus also decreases the associated efficiency.

In contrast, when the wings' inclination changed when they were in proximity
such as when the minimum spacing was reached at $ft=0.826$, it was shown that
induced velocities resulting from 3D effects were significantly reduced. In such
a configuration, the pressure distribution resulting from a beneficial
ground-effect-like phenomenon created by the two wings resulted in an upwash
relative to the upper wing on its internal surface (thus a downwash for the
lower wing). However, due to the proximity of the wing tip vortices resulting
from the clap-and-fling mechanism, the resulting induced velocities of each
vortex in opposite directions cancel each other out between the wings. As a
result, the effective angle of attack was not strongly impaired, which led to
performances similar to the 2D prediction. Having wings in proximity was thus
beneficial to mitigate some adverse 3D effects, further illustrating the
benefits of the clap-and-fling phenomenon employed in a three-dimensional
configuration.

Moreover, it was shown through Q-criterion visualizations as well as spanwise
propulsive and power distributions that the stability of the ejected spanwise
vortices from the boundary layers are strongly dependent on the wings' aspect
ratio. Indeed, for wings with low normalized spans, the proximity of the strong
wing tip vortices caused by the three-dimensional physics flattened the vortices
adjacent to the boundary layers oriented in the spanwise direction.  For wings
with an increasing aspect ratio, the spanwise vortices stayed correlated for a
larger distance near the middle of the wings with complex vortical structures
around the latter resulting from instabilities caused by wing-tip vortices.
Furthermore, this change in the behavior of the flow greatly affected the
propulsive and power spanwise distribution with a plateau that arises near the
midsection of the wing, especially when 3D effects are maximized in instances
such as when the wings are further apart. In contrast, those negative
phenomenons do not occur for wings with high normalized spans such as $b^*=4$
and $b^*=8$ where the spanwise distribution of vortical structure maintained its
integrity and thus yielded better performances.

The 3D simulations conducted utilized a somewhat complex motion including
heaving, pitching as well as deviation which was proven in a previous study
\citep{Papillon_clap_fling_2024} to amplify the unsteady effects and thus
increased the performances of a system subjected to the clap-and-fling
mechanism. By considering the same in-plane kinematics in a three-dimensional
space, aerodynamic losses were directly evaluated, which gave insight into the
behavior of a system using the clap-and-fling mechanism as a means of
propulsion. The knowledge of how 3D effects decrease aerodynamic performances
and how they can be mitigated using long aspect ratio wings that flap in
proximity brings a better understanding of flapping-wing flight and its
potential applications.

\clearpage
\bibliographystyle{plainnat} 
\bibliography{Article_clap_fling_3D}
 
\end{document}